%% file: main.tex
\documentclass[12pt,twocolumn]{openjournal}
\usepackage{xcolor}
\usepackage{textgreek}
\usepackage[utf8]{inputenc}
\usepackage[english]{babel}
\usepackage{hyperref}
\hypersetup{
    colorlinks=true,
    linkcolor=blue,
    filecolor=blue,      
    urlcolor=blue,
    citecolor=blue,
}
\usepackage{color,colortbl}
\usepackage{tensind}
\tensordelimiter{?}
\DeclareGraphicsExtensions{.bmp,.png,.jpg,.pdf}
\usepackage{verbatim}
\usepackage[normalem]{ulem}
\usepackage{orcidlink}
\usepackage{soul}
\usepackage{multirow}

\newcommand{\orcidauthor}[3]{\author{\href{http://orcid.org/#1}{#2$^{#3}$}}}

\graphicspath{ {./figs/} }

\begin{document}
\title{DB-Bench: Benchmarking Deblenders for LSST DESC Using the Blending ToolKit\vspace{-4em}}

%\author{Aidan Berres\orcidlink{0000-0000-0000-0000}}
%\email{aberres2@illinois.edu}
%\affiliation{Department of Astronomy, University of Illinois at Urbana-Champaign, Urbana, IL 61801, USA}

%\author{Grant Merz\orcidlink{0009-0005-7923-054X}}
%\email{gmerz3@illinois.edu}
%\affiliation{Department of Astronomy, University of Illinois at Urbana-Champaign, Urbana, IL 61801, USA}

%\author{Xin Liu\orcidlink{0000-0003-0049-5210}}
%%\email{xinliuxl@illinois.edu}
%\affiliation{Department of Astronomy, University of Illinois at Urbana-Champaign, Urbana, IL 61801, USA}
%\affiliation{National Center for Supercomputing Applications}
% Cécile Roucelle : Université Paris Cité, CNRS, AstroParticule et Cosmologie, F-75013 Paris, France

% Éric Aubourg : Université Paris Cité, CNRS, CEA, AstroParticule et Cosmologie, F-75013 Paris, France

\orcidauthor{0000-0002-5010-441X}{Aidan Berres}{1,*}
\orcidauthor{0009-0005-7923-054X}{Grant Merz}{1,*}
\orcidauthor{0000-0003-0049-5210}{Xin Liu}{1,2,3}
\orcidauthor{0000-0002-5592-023X}{\'{E}ric Aubourg}{4}
\orcidauthor{0000-0002-9641-4552}{C\'{e}cile Roucelle}{5}
\author{the LSST Dark Energy Science Collaboration}

\thanks{$^*$Corresponding Authors: Aidan Berres (\href{mailto:aberres2@illinois.edu}{aberres2@illinois.edu}), Grant Merz (\href{mailto:gmerz3@illinois.edu}{gmerz3@illinois.edu}).}

% List of institutions
\affiliation{$^{1}$Department of Astronomy, University of Illinois at Urbana-Champaign, 1002 West Green Street, Urbana, IL 61801, USA}
\affiliation{$^{2}$National Center for Supercomputing Applications, University of Illinois at Urbana-Champaign, 1205 West Clark Street, Urbana, IL 61801, USA}
\affiliation{$^{3}$Center for Artificial Intelligence Innovation, University of Illinois at Urbana-Champaign, 1205 West Clark Street, Urbana, IL 61801, USA}
\affiliation{$^{4}$Universit\'{e} Paris Cit\'{e}, CNRS, CEA, AstroParticule et Cosmologie, F-75013 Paris, France}
\affiliation{$^{5}$Universit\'{e} Paris Cit\'{e}, CNRS, AstroParticule et Cosmologie, F-75013 Paris, France}

\begin{abstract}
    Blending will be a major source of systematic uncertainty in downstream science analyses of LSST data. We benchmark the performance of several deblenders, leveraging the BlendingToolKit (BTK) to perform rigorous, end-to-end testing. This benchmark incorporates key deblending algorithms, including SourceExtractor, SCARLET, and DeepDISC, with the goal of comparing their effectiveness in handling blended galaxy images from LSST/Rubin simulations. A key focus is characterizing algorithm performance in the regime of unrecognized blends, where multiple galaxies are misidentified as a single object, as these cases introduce systematic biases that propagate into downstream cosmological analyses for galaxy surveys. By utilizing BTK's ability to create customized, reproducible blends, we systematically test these deblenders against different blending conditions, such as source separation and brightness. The toolkit's standardized evaluation metrics, including detection precision, segmentation accuracy, and source reconstruction, are comprehensive assessments of each algorithm's strengths and limitations. Each deblender has performance caveats that may impact their true performance in real survey conditions. We find that SCARLET has high segmentation and reconstruction performance, whereas DeepDISC has strong detection recall for faint and low-SNR sources, and SourceExtractor has accurate peak finding abilities but low segmentation and reconstruction performance. This benchmark provides valuable insights into the performance of existing deblenders and highlights areas for future development.

\end{abstract}
%Our goal is for this initiative to evolve into a DESC data challenge, encouraging broader community participation.
% Additionally, this project incorporates challenging cases such as unrecognized blends, pushing the deblenders to their limits. 
% These unrecognized blends are a major source of error for many downstream analyses using galaxies. BTK provides a robust simulation framework based on GalSim, capable of generating realistic galaxy blends under varying observing conditions. 
% Write your keywords here

\maketitle

\section{Introduction}\label{sec:intro}

The Legacy Survey of Space and Time (LSST) at the Vera C. Rubin Observatory \citep{lsst_2019} will deliver an unprecedented census of the optical sky, enabling precision measurements of billions of galaxies for cosmology and astrophysics. A dominant and increasingly recognized source of systematic uncertainty in LSST analyses is galaxy blending, in which overlapping light profiles bias object detection, photometry, morphology, and shape measurements \citep{melchior_2021}. The density of individually resolved sources will drastically increase as surveys like LSST probe fainter sources and increase their sensitivity, however this increases the amount of overlapping sources being resolved as well. Studies from the Hyper Suprime-Cam Subaru Strategic Program (HSC SSP) \citep{HSC} found that at their (i.e. similar to LSST) depths $58 \%$ of detected galaxies were identified as blended \citep{hsc_pipeline}. Previous surveys have found blending to have a serious impact on galaxy measurements. The Dark Energy Survey Y1 (DES) \citep{DES} found that $\sim 30 \%$ of galaxies observed for weak lensing shear measurements would have to be removed due to having close neighbors to reduce shear bias \citep{samuroff_2018}. Further studies on the impact of blending on ellipticity measurements found that blended galaxies increase the overall shear noise by $14 \%$ \citep{dawson_16}. These uncertainties will propagate directly into downstream science analyses including photometric redshifts \citep{des_yr1_photoz} and cosmic shear \citep{Nourbakhsh_22,sanchez_overlap_cosmic_shear}. As source densities increase toward faint magnitudes and in crowded environments, robust deblending is no longer a secondary technical step but a foundational requirement for reliable LSST science.

Traditional deblending techniques, such as the widely used SourceExtractor \citep{sextractor}, excel at separating moderately blended sources by applying morphological filters but struggle in complex or crowded fields \citep{molino_2017,melchior_2021}. More recent algorithms leverage multi-band modeling and non-parametric constrained optimization. Notable examples include MuSCADeT \citep{Joseph2016}, which introduced multi-band morpho-spectral component analysis, and its successor SCARLET \citep{scarlet}, which generalizes this framework to allow arbitrary source constraints and is currently deployed in the HSC and LSST pipelines. Deep learning-based deblenders have subsequently demonstrated strong reconstruction performance in LSST-like simulations, including VAE-based approaches \citep{vae_reconstructions}, GAN-based methods \citep{Reiman2019,Hemmati2022}, hybrid physical-deep learning models \citep{Lanusse2019}, residual dense networks \citep{Wang2022}, score-matching priors extending SCARLET \citep{Sampson2024}, MADNESS \citep{madness}, and BLISS \citep{bliss,Mendoza2026}. For a comprehensive overview of blending challenges and mitigation strategies, see \citet{melchior_2021,Xu2024}.

\begin{figure}
    \centering
    \includegraphics[width=1\linewidth]{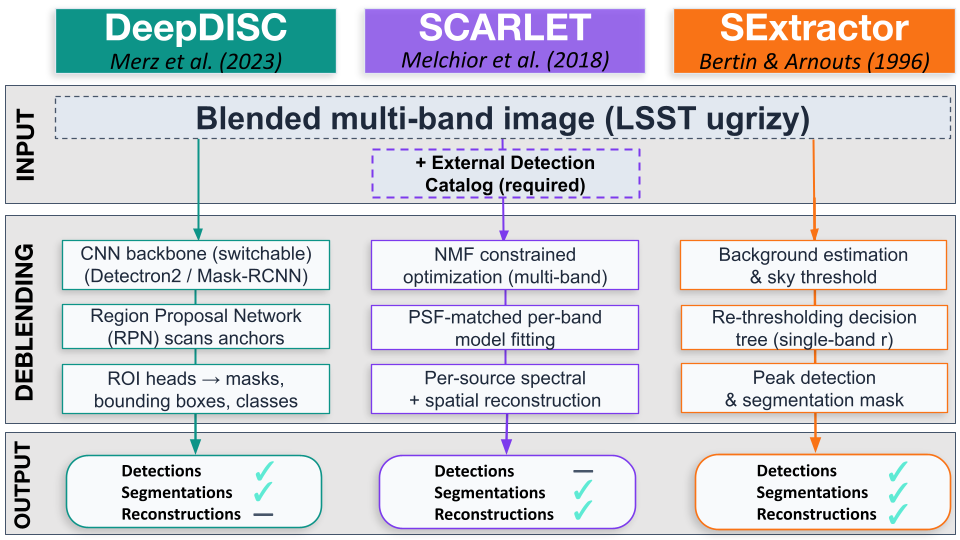}
    \caption{Comparison of the deblending frameworks benchmarked in this work. DeepDISC \citep{deepdisc} and SourceExtractor \citep{sextractor} both detect sources directly from the input image and can therefore be compared at the full pipeline level; SCARLET \citep{scarlet} requires an externally provided detection catalog as input and is therefore evaluated only at the deblending component level. The three deblenders share a common blended image input but differ in their internal processing and in the output data products they produce. Checkmarks indicate which data products are available from each algorithm; note that DeepDISC does not produce reconstructed galaxy images, and SCARLET does not perform source detection.}
    \label{fig:deblenders}
\end{figure}

Complementing these methods is DeepDISC \citep[][see also \citealt{Burke2019}]{deepdisc,deepdisc_photoz,Merz2026}, a deep learning-based approach tailored to LSST’s specific needs and capable of computationally efficient deblending and classification. 
%This novel addition to the BTK allows the first direct comparison of a deep-learning based deblender with more traditional algorithmic methods.
These techniques largely differ at the pipeline level, with SCARLET relying on a detection catalog of sources whereas SourceExtractor and DeepDISC can detect sources directly from an image, thus not requiring an independent detection pipeline. DeepDISC uses several Region Proposal Networks (RPNs) and Region of Interest (ROI) heads to identify and segment multiple sources in an image. This can allow for a direct comparison of detection performance at the pipeline level for SourceExtractor and DeepDISC (\autoref{fig:deblenders}). Comparisons with SCARLET are done at the specific deblending component level. These comparisons are made with the understanding that detection performance directly impacts the output of a deblender.
%BLISS, a Bayesian approach, incorporates priors on galaxy morphology to iteratively fit models to blended sources, showing promise in handling faint and irregular galaxies \citep{bliss}.

Despite the central role of deblending, the community lacks a systematic, end-to-end benchmark of deblending algorithms under controlled yet realistic LSST conditions. Specifically, this is missing in regimes where blends are unrecognized, i.e., where multiple galaxies are detected as a single object. In this work, we address this gap by comparing the performance of two traditional and one deep-learning based deblenders: SourceExtractor, SCARLET, and DeepDISC using the BlendingToolKit (BTK) \citep{btk}. The version of the BTK that is used for this study was modified from its original version with the addition of DeepDISC\footnote{\url{https://github.com/berres2002/BlendingToolKit}}. Our evaluation uses reproducible LSST-like galaxy blends spanning signal-to-noise, separation, and source density. By evaluating detection, segmentation, and reconstruction performance across both simple blends and large, crowded scenes, the regimes of validity and failure modes of each method can be identified. The results are intended not to crown a single ``best'' deblender, but to inform pipeline design choices for LSST and to provide a quantitative foundation for future LSST Dark Energy Science Collaboration (DESC) deblending challenges.

In this study, we aim to conduct a rigorous benchmarking of these deblenders, utilizing BTK as a standardized, end-to-end testing framework. BTK, which employs GalSim \citep{galsim} to generate realistic galaxy blends, is a powerful tool for controlled experimentation across different blending conditions. This includes variable brightness, galaxy separation, and observing conditions. The main focus of this benchmark is to map failure modes for each deblender and comparing their performance between regimes. Our novel analysis simulates blended scenes at LSST-like scales while sampling from the LSST CosmoDC2 simulated galaxy catalog \citep{cosmoDC2}. By leveraging BTK’s capabilities, it is possible to systematically simulate diverse blending scenarios that reflect LSST’s observing challenges. An essential focus will be on ``unrecognized blends'', scenarios where the blending is so severe that multiple sources are mistaken for a single object. These cases, which remain largely under-explored, introduce biases that propagate through science pipelines and are particularly relevant for weak lensing studies, where accurate shape measurement is critical \citep{sheldon_12,hoekstra_17}. 
% This allows for a nuanced evaluation of each algorithm’s performance. 
% Additionally, BTK provides a suite of standardized metrics—such as detection precision, segmentation accuracy, and shape reconstruction accuracy—that enable quantitative assessments of each deblender's strengths and limitations.
Understanding the boundaries of current deblending techniques and identifying areas of future development require systematically testing each algorithm against these complex cases.

% The main focus of this benchmark is to map failure modes for each deblender and comparing their performance between regimes. An essential focus will be on ``unrecognized blends'', scenarios where the blending is so severe that multiple sources are mistaken for a single object. These cases, which remain largely under-explored, introduce biases that propagate through science pipelines and are particularly relevant for weak lensing studies, where accurate shape measurement is critical \citep{sheldon_12,hoekstra_17}. These blends could also affect Photometric Redshift (Photo-z) measurements of galaxies which require accurate photometry \citep{liang_adari_unrec_blends}. \texthl{Understanding the boundaries of current deblending techniques and identifying areas of future development require systematically testing each algorithm against these complex cases.}

% Systematically testing each algorithm against these complex cases, we aim to understand the boundaries of current deblending techniques and identify areas requiring further development.

\begin{figure}
\centering
\includegraphics[width=1\columnwidth]{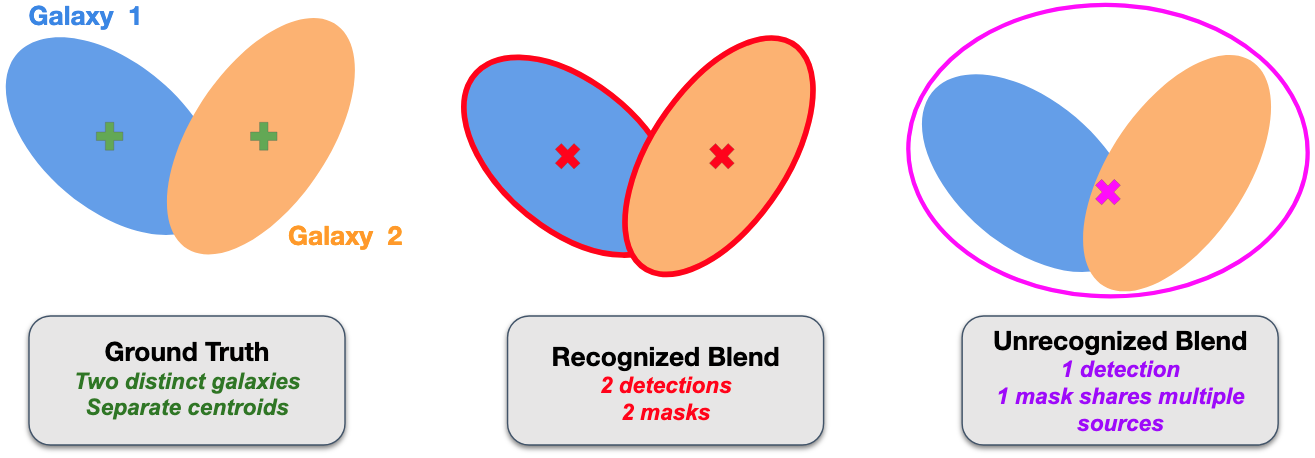}
\caption{Schematic illustration of the three outcomes relevant to galaxy deblending. \textit{Panel A (Ground Truth):} Two physically distinct galaxies overlapping along the line of sight, with independently known centroids and flux profiles. \textit{Panel B (Recognized Blend):} A deblender correctly identifies two distinct sources, returning separate segmentation maps and centroid estimates. Downstream measurements remain imperfect but are individually attributed to each galaxy. \textit{Panel C (Unrecognized Blend):} A deblender returns a single detection encompassing both galaxies. The combined object is treated as a single source throughout the science pipeline, introducing systematic biases in flux, shape, photometric redshift, and weak lensing shear measurements. Unrecognized blends are the dominant failure mode of SourceExtractor in crowded and highly blended scenes, and occur at lower rates in DeepDISC (see Section \ref{sec:det_results}).}
\label{fig:unrecognized_blend}
\end{figure}

Ultimately, we envision this benchmarking effort as a foundation for a DESC data challenge, inviting broader community participation and fostering the development of improved deblending techniques suited to the LSST era. There has been broad community adoption of Artificial Intelligence (AI) and Machine Learning (ML) based analysis tools in the DESC community (See \cite{desc_ai_2026_WHITE_PAPER}). This has led to the development of many ML based deblenders specifically for LSST observations. This data challenge will allow current methods to be evaluated and promote the development of new ML based deblenders. Such challenges will be instrumental in aligning the community toward standardized evaluation frameworks and performance metrics, thereby advancing the reliability and robustness of galaxy deblending techniques essential for LSST and similar large-scale astronomical surveys.

This paper presents three main innovations: 
\begin{enumerate}
    \item First, a benchmarking of representative deblending algorithms: SourceExtractor, SCARLET, and DeepDISC, using BTK to systematically test performance under realistic LSST/Rubin simulations.
    \item Second, we introduce challenging scenarios, including unrecognized blends (\autoref{fig:unrecognized_blend}), providing insights into each algorithm's robustness and adaptability.
    \item Third, we offer a rigorous framework that may serve as the foundation for a DESC data challenge, encouraging broader community involvement and fostering advancements in deblending methodologies.
\end{enumerate}

The structure of this publication is as follows: Section \ref{sec:eval} provides a detailed overview of the BlendingToolKit framework and the specific simulation configurations used. Section \ref{sec:deblend} describes the deblending algorithms benchmarked in this study, with emphasis on their unique methodologies and limitations. Section \ref{sec:results} presents the evaluation metrics and results of our benchmarking, including performance comparisons across diverse blending scenarios. In Section \ref{sec:discussion}, we analyze the implications of these findings for LSST science and propose areas for future deblender development. Finally, Section \ref{sec:concl} summarizes our findings and discusses potential future projects. 
%The section outlines our vision for expanding this benchmarking project into a DESC data challenge, aiming to advance the community’s collective efforts in deblending techniques for large astronomical surveys.

\section{Evaluation Framework}\label{sec:eval}

\subsection{Blending ToolKit}\label{sec:btk}

\begin{figure}
\centering
\includegraphics[width=0.5\textwidth]{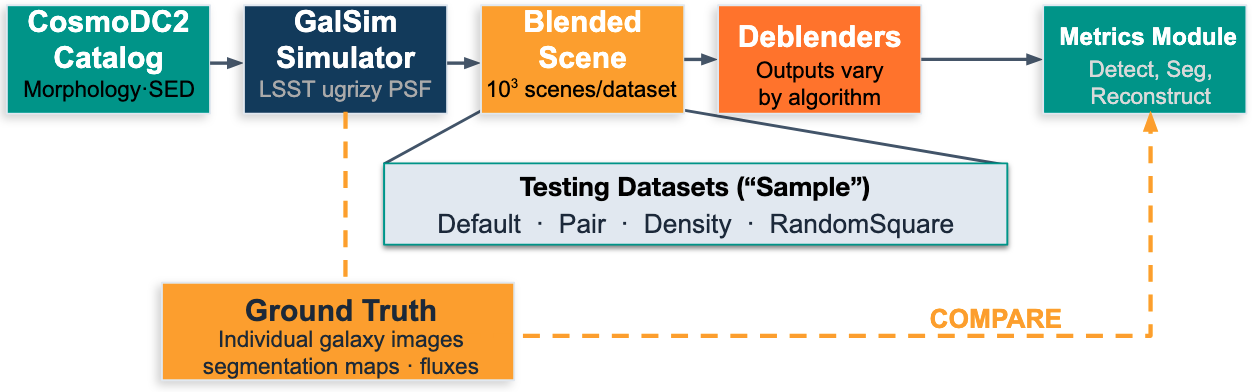}
\caption{Schematic of the BlendingToolKit (BTK) end-to-end benchmarking pipeline used in this work. Galaxy properties are sampled from the cosmoDC2 simulated catalog \citep{cosmoDC2} and passed to GalSim \citep{galsim}, which simultaneously produces blended scene images (one per LSST $ugrizy$ passband) and noiseless individual galaxy images used as ground truth. Four sampling strategies generate distinct datasets: Default (highly blended), Pair (separated), Density (galaxy cluster), and RandomSquare (survey-scale), each consisting of 1000 scenes. Blended images are passed to each deblender module, and the resulting outputs (detection catalogs, segmentation maps, and reconstructions) are compared against ground truth by the metrics module.}
\label{fig:btk_workflow}
\end{figure}

The Blending ToolKit (BTK) \citep{btk} is a framework for testing the performance of various deblending software. It combines galaxy image simulations and the output of deblenders to create an end-to-end testing platform where the generated ``ground truth'' data is compared with the final deblended results with various testing metrics (\autoref{fig:btk_workflow}). BTK contains 3 main modules: generation, deblending, and metrics. The generation process begins by sampling galaxy catalogs with properties like morphology measurements and spatial brightness data to generate realistic blended images. The generation module uses GalSim \citep{galsim} to generate the test data that is given to the deblenders.
% These sampling methods assemble the simulated ground truth detection catlog for each blended image. 
The choice of sampling method determines the amount and attributes of the galaxies drawn for each blended scene. The sampling methods and source catalog used in this work are discussed in Section \ref{sec:dataset}. 
% This process creates both individual images of each galaxy and the combined blended scene. 

These generated images are then given to the deblending module which can produce detection catalogs (source R.A. and Dec), segmentation maps (pixel-level masks of each galaxy in an image), and reconstructions (individual galaxies redrawn on their own image). The list of deblenders used in this implementation of BTK is discussed in Section \ref{sec:deblend}. 

The metrics module then takes these deblended outputs and evaluates them using tailored functions for each data product. The evaluation framework and testing metrics used in this work are described in Section \ref{sec:metrics}. This work aims to promote the future use of and contributions to BTK specifically with the addition of new deblender pipelines (e.g. MADNESS \citep{madness}, BLISS \citep{bliss}, and Morpheus \citep{Hausen2020}) and metrics, which is discussed in Sections \ref{sec:expand_btk} and \ref{sec:data_challenge}.

\begin{figure*}[ht!]
    \centering
    \includegraphics[width=1\linewidth]{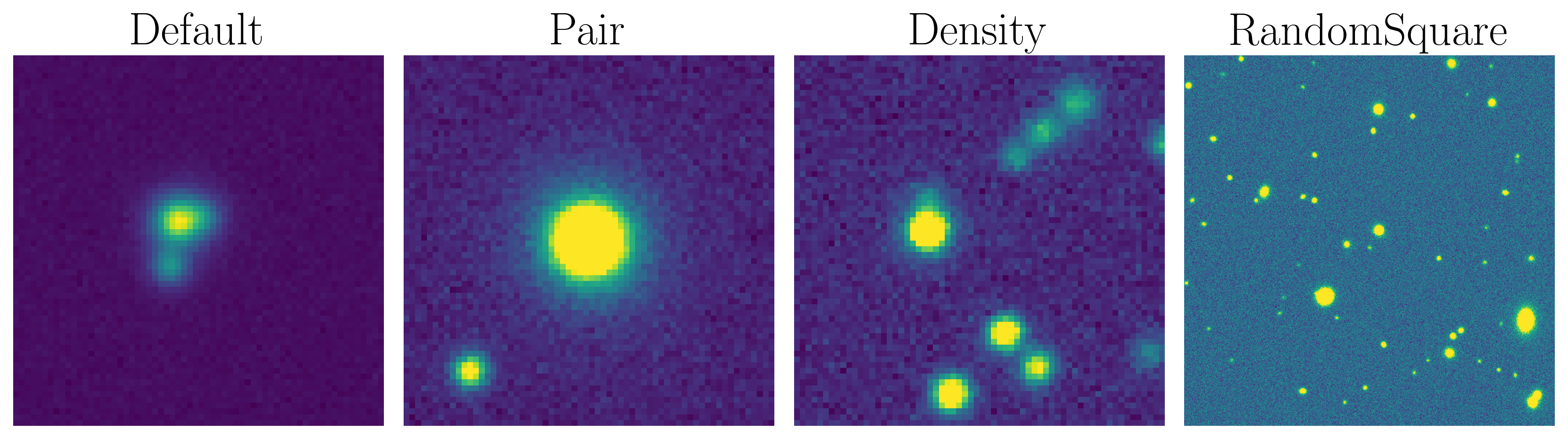}
    \caption{Sample images from each dataset, summed over every LSST band (\textit{ugrizy}). Each dataset represents different blending situations which are labeled in parenthesis and used the same source catalog with a different sampler: (\textit{from left to right}) a) Default Sampling Function (Highly Blended) b) Pair Sampling Function (No Blending), c) Density Sampling Function (Galaxy Cluster), d) RandomSquare Sampling Function with higher resolution $512 \times 512$ pixel image (Survey Image).}
    \label{fig:datas}
\end{figure*}

\subsection{Generated Test Dataset}\label{sec:dataset}

The dataset of blended images made for this benchmark are sampled from the \texttt{cosmoDC2} simulated galaxy catalog \citep{cosmoDC2,dc2_creation}. \texttt{cosmoDC2} (DC2) was created to allow various LSST DESC working groups to prepare and test data analysis pipelines before LSST became active. DC2 only captures a small portion of the full LSST observational footprint at only 300 $\text{deg}^2$. The data was generated using N-body particle simulations that produce halos and lightcones based on the observer location. This simulation is run at different redshift depths, starting at $z \sim 3$ and ending at $z=0$. Resultant halos are assigned galaxies using the UniverseMachine synthetic galaxy catalog \citep{universemachine} by matching similar halo masses. Then weak lensing properties are assigned to each galaxy using the particle cone simulations. Galaxy properties like morphology, photometry, and SEDs are assigned based on matching from UniverseMachine and Galacticus models.

% \setcounter{footnote}{0}
% \addtocounter{footnote}{-1}

We use a 1 $\text{deg}^2$ subset of the DC2 catalog to simulate data for benchmarking. The generation workflow begins by sampling galaxies from the catalog using sampling functions provided by BTK. This creates a sample catalog for each blended scene. These sampled catalogs are then given to a blend generator which employs GalSim to draw galaxies on an image given the properties from the simulated catalog. These properties include multi-band magnitudes, AGN activity, and morphological properties like bulge and disk sizes. Furthermore, properties of the observing instrument are also applied including pixel scales, per-band exposure time, and a simulated Point Spread Function (PSF). Every image is generated based on the LSST/Rubin observatory's observing conditions and instrumentation, with one image for each \textit{ugrizy} passband. The datasets reflect the observing limitations of 10 years coadded depth with LSST using a Poisson simulated noise and simulated Point Spread Function (PSF). This coadded depth is simulated from the amount of noise expected for each passband over 10 years using the \texttt{surveycodex}\footnote{\url{https://github.com/LSSTDESC/surveycodex}} package. The magnitude range for each source was chosen to be $24.5 \ge m_r \ge 18$, with an upper limit chosen to avoid saturation and limit larger sources that would cover small blending scenes. A galaxy in a blended scene is drawn individually on an image and then combined into a single blended image. The individually drawn galaxy images are used as detailed ground truth comparisons for deblender predictions like segmentation maps and reconstructions. A full test data set is composed of 1,000 unique blended scenes in every LSST filter.

% A standard sample blended image is a 64 pixel$^2$ image at a pixel scale of 0.2 arcsec per pixel with N max number of galaxies. 

% For the Default Sampling function, $24.5 \ge m_r \ge 18$ \citep{lsst_2019}. Furthermore, the maximum shift between blended galaxies was set to 1 arcseconds. A full test data set is composed of 1,000 unique blended scenes in every LSST filter.

% The data is generated per batch with a set amount of images in each batch. This batch is then handed to each deblender and deblended results are given per blend batch. 
This benchmark uses four different sampling functions that generate distinct datasets each with different blending situations. 'Pair Sampling' as a control, 'Default Sampling' to test for potential unrecognized blends, 'Density Sampling' for dense galaxy clusters, and 'RandomSquare Sampling' for larger survey-like scenes. Examples of these sampling function outputs are shown in Figure \ref{fig:datas} and specific attributes in Table \ref{tab:samp_attribs}.

The first dataset generated used the Default Sampling function. This function allows for the sampling of a specified number of galaxies to be drawn at a fixed distance apart. This dataset consists of 3 galaxies per image with the maximum centroid offset between blended galaxies set to 1 arcseconds. This is to emulate only highly blended scenes as the sources in this set always overlap. The resolution of this set is 64 pixel$^2$ at a pixel scale of 0.2 arcsec per pixel.

% Other sampling methods used in this work \texthl{are} Pair Sampling, Density Sampling\texthl{, and RandomSquare Sampling}. These methods provide random and controlled situations commonly found in real observations. 
Pair Sampling outputs a pair of galaxies with a specified distance apart where one galaxy is brighter than the other, within specified magnitude limits. The separation used in this dataset is a 5 arcsecond separation, with the same magnitude limits imposed in the Default Sample. This separation ensures that this dataset consists of separated sources where there is minimal blending by construction. This is to test deblender performance on sources with minimal blending. 

Density Sampling outputs a set number of galaxies in a random field within a set density of photon counts per square arcminute. This emulates galaxy clusters with many sources close to each other. We use 10 randomly placed galaxies of varying brightness (within the same magnitude limits as the Default Sample) within an image. The centroid location of each galaxy is within the image coordinates, however extended light profiles may be truncated by the edge of the image cutout. 
%Varying the parameters of these sampling functions allows one to gauge the performance of each deblender with different controlled situations. 

RandomSquare Sampling uses the pre-existing coordinates found in the DC2 simulated galaxy catalog to sample every galaxy found within a randomly located 102.4 $\times$ 102.4 arcsec square. This dataset generates larger images to understand how each deblender's performance can scale to much larger ``survey-like" scenes.
% A dataset with larger images is also generated to understand how each deblender's performance can scale to much larger \texthl{``survey-like" scenes}. 
% \texthl{These images are} generated with a resolution of 512 pixel$^2$ using the BTK's \texttt{RandomSquareSampling} sampler function. This function uses the pre-existing coordinates found in the DC2 simulated galaxy catalog to sample every galaxy found within a randomly located 102.4 $\times$ 102.4 arcsec square. 
Unlike previous sampling methods, this sampler preserves the original coordinates of each galaxy when drawn in a blended scene, instead of sampling galaxies and drawing them at a specified location. This provides a more realistic survey dataset, with each blended image containing up to 200 galaxies. A full set of 1,000 RandomSquare images results in GalSim drawing $\sim 79,000$ individual galaxies. %\textcolor{red}{How many galaxies in the entire set?}

% m{2.2cm}|c|c|c
\renewcommand{\arraystretch}{2.5}
\begin{table*}
    \centering
    \begin{tabular}{m{2.5cm}m{2.1cm}m{2.5cm}m{2.5cm}m{1.5cm}m{3cm}}
        \hline
        \textbf{Sample} & \textbf{Magnitude Range} & \textbf{Max Separation} (arcsec) & \textbf{Galaxies per Image} & \textbf{Resolution} (pixel$^2$)& \textbf{Target Case}\\
        \hline
        Default & $24.5 \ge m_r \ge 18$ & 1 & 3 & 64& Unrecognized blends. High blending.\\
        Pair & " & 5 & 2 & "& Control. No blending.\\
        Density & " & N/A & 10 & "& Galaxy Clusters.\\
        RandomSquare & "& N/A & $\leq200$ & 512& Sky Survey-like.\\\hline
    \end{tabular}
    \caption{Generated Test Dataset Sample Attributes.}%" represents a repeated entry from above.
    \label{tab:samp_attribs}
\end{table*}

\subsection{Testing Metrics} \label{sec:metrics}

Our testing framework uses metrics included in the standard BTK installation. The framework tests each deblender's performance on 3 categories, \textit{Detection}, \textit{Segmentation}, and \textit{Reconstruction}. Each metric category has its own method of measurement and requires different forms of deblender output. To measure the \textit{Detection} metric, the locations of the galaxies in the detection catalog is compared with the individual simulated galaxies in the ground truth source catalog. 
% This output catalog acts as a source detection catalog which requires the deblender to detect the sources in the input image. 
Deblenders like DeepDISC (See \ref{sec:deepdisc}) and SourceExtractor (See \ref{sec:sextractor}) detect sources when deblending. The \textit{Segmentation} metric measures the accuracy of the identifying the pixel area around each galaxy. This is compared to the maps generated by placing a sky background noise threshold on the individual galaxy images generated by BTK and placing a mask over this threshold. Measuring \textit{Reconstruction} requires the deblender to reconstruct the individual galaxy images from the input blended image. These reconstructions are compared with the ground truth unblended galaxies generated by GalSim. %See Table \ref{tab:metrics} for each metrics description within the output of this benchmark.

% Detection matters for spatial correlation analyses like 3x2pt analysis
% Segmentation is important for weak lensing measurments that require accurate shape measurements
% Segmentation is also important for Photo-zs
% Reconstruction is important for Photo-zs since isolation and proper photometry is important

% Detection Metrics
\textit{Detection} performance is evaluated with ``Precision" and ``Recall" metrics.
% The evaluation method used to measure \textit{Detection} is the ``Precision" and ``Recall" metrics. 
% These metrics evaluate the quality and the accuracy of detected sources by dividing the number of matched detections by the number of total detections for precision and by the number of true sources for recall. 
Precision indicates the percentage of accurate predictions the deblender makes out of the total number of predictions. Likewise, recall values indicates the percentage of true sources that were recovered by the deblenders. The matching is conducted by BTK's \texttt{PixelHungarianMatcher} which uses the pixel coordinates of galaxy centroids. The minimum threshold distance for a detection to be matched to the ground truth is set at 5 pixels or $1''$. We note that for the Default sample, where source centroids are drawn within ${\leq} 1''$ of each other, this matching radius is comparable to the source separation itself. Cross-matches between adjacent sources are therefore possible in this regime, and detection recall for the Default sample should be interpreted as a conservative lower bound on true deblending performance. This separation is chosen in order to identify the rate of unrecognized blends for extremely close sources. Locating galaxies in a blended scene directly impacts spatial correlation analyses like 3x2pt analyses mapping dark energy and weak lensing \citep{3x2pt_desi}. Furthermore, understanding the rate of ``detection" vs. ``non-detection" with different blended scenes provides an insight to where unrecognized blends could be most prevalent.

% Segmentation metrics
The \textit{Segmentation} metric is measured using ``Intersection-over-Union (IoU)" metric. The IoU metric measures the overlap between the output segmentation map and the ``ground truth" map determined by BTK. The IoU is the ratio of the pixel area shared by the ground truth segmentation ($A$) and the output segmentation maps ($B$), $A \cap B$, and the total area covered by both maps, $A \cup B$, thus $\text{IoU} = A \cap B / A\cup B$. An ideal IoU value would be close to 1 as the intersection area and total areas would be close to identical, whereas an IoU value closer to zero would indicate little overlap ($A \cap B$) between segmentation maps. These maps allow researchers to understand the potential shape of sources in a blended scene which can impact Weak Lensing (WL) studies that require accurate shapes to measure cosmic shear and dark matter \citep{dawson_16, Nourbakhsh_22}.

 It should be noted that SCARLET does not natively output segmentation maps; instead, its segmentation maps are derived from reconstructed galaxy images using the same BTK thresholding procedure used to construct ground truth. This means SCARLET's IoU values are not directly comparable to those of DeepDISC and SourceExtractor, which produce independent pixel-level segmentations, and should be interpreted with this caveat in mind.

% of these two maps and the non-intersecting area. 

% Reconstruction metrics
This benchmark framework uses the ``Structural Similarity Index (SSIM)" \citep{ssim} and Mean Squared Error (MSE) as its \textit{Reconstruction} metrics. The SSIM is a pixel-level analysis that specifically compares each image's (\textit{l}), contrast (\textit{c}), and structure (\textit{s}). The luminance is calculated from the mean intensity, the contrast is determined by the standard deviation of each images pixels minus the mean intensity, and the structure is determined using pixel-level values with the mean intensity subtracted and then divided by its standard deviation. These values are then multiplied together to provide the metric value.  An ideal SSIM value for a reconstructed source would be 1. SSIM is used as one of the Reconstruction metrics as it quantifies the degradation (e.g. changes) in structure between two images. Since deblenders are fully replicating sources, preserving structure like shape and light profile are imperative. Reconstructing individual sources from a blended scene can improve individual photometric measurements \citep{vae_reconstructions,scarlet}.  The Mean Squared Error (MSE) is also used as a reconstruction metric to provide a total pixel-level error between the output and the ground truth. 

\section{Deblending Algorithms}\label{sec:deblend}

% Galaxy deblenders are pieces of software 
BTK includes several deblenders for testing and allows for more deblenders to be implemented into its framework. Current deblending methods already incorporated into BTK include SourceExtractor Python (SEP), a SourceExtractor Python wrapper, and SCARLET, an advanced catalog-based algorithm using matrix factorizations and symmetries to reconstruct galaxies from a blend. For this work, DeepDISC was included into BTK to compare its results with these two deblenders.

\subsection{DeepDISC}\label{sec:deepdisc}

DeepDISC \citep{deepdisc} is an instance segmentation software that is based on a Region-based Convolutional Neural Network (RCNN) from Facebook AI's Detectron2 model \citep{wu2019detectron2}. It specifically uses a vision transformer or Convolutional Neural Network (CNN) feature extractor backbone which is sampled by a Region Proposal Network (RPN) to compile a list of potential targets. These sampled targets are then passed to the Region of Interest (ROI) heads in which they are given annotations like segmentation masks and bounding box coordinates. DeepDISC can detect sources from input images and does not require a detection catalog as input. It was initially used with HyperSuprime Cam (HSC) data \citep{HSC} and has shown increasing performance on detecting and deblending faint galaxies in a given sample of blended images. For more information on the architecture and applications of DeepDISC please see \cite{deepdisc}. Furthermore, DeepDISC has been expanded to identify photometric redshifts of each deblended galaxy by adding new annotations to its ROI heads (See \citet{deepdisc_photoz}). 
% The model weights used in this benchmark were trained with data generated from the BTK's blend generator. 

Unlike deblenders that do not use machine learning, DeepDISC's performance is dependent on the data it is trained on.
Since DeepDISC is a machine learning based model, its performance is dependent on its training data. It has been found that its performance increases as the model becomes more generalized given more randomized and realistic training data.
For this benchmark a CNN feature extractor backbone in DeepDISC's architecture is used. Two distinct DeepDISC models, trained on BTK-generated images are used in this benchmark. BTK and Galsim generate both ground truth information and blended scenes which we utilize to train and validate our DeepDISC models. For each model the training and test split were 2000 training images with 200 validation images. 
% \texthl{An initial model trained only on Default sampled images was used to test the implementation of DeepDISC in BTK. This model is not used in the current benchmark.}
The first model, using only Density sampled images to replicate more varied and realistic training data, is utilized in DeepDISC's benchmark for the Default, Pair, and Density data sets.
For the larger RandomSquare sample dataset, a second model trained on RandomSquare sampled data is used. This is done to evaluate the performance of a model trained exclusively on survey data. 
% \texthl{Performance increases as the model becomes more generalized given more randomized and realistic training data. The more specific the training set is (i.e. the first Default sample set) less performance gains are found when it is used on data that does not conform to the set.}
It is of note that the input image size for DeepDISC can be flexible and is not dependent on the image size used to train the model.
We further note that because DeepDISC's models are trained on images generated by the same BTK/GalSim pipeline used for evaluation, its performance reflects an idealized in-distribution scenario. Performance on actual Rubin LSST data, 
% which will differ from GalSim simulations in PSF modeling, detector artifacts, and background complexity, 
may differ from the results reported here.

% DeepDISC can detect sources from input images and does not require a detection catalog as input. This allows for the measurement of DeepDISC's detection accuracy within this benchmarking framework. Furthermore, as DeepDISC outputs segmentation maps its segmentation accuracy can also be determined. It is also possible to measure its reconstruction ability by using the segmentation maps to mask out each source to create individual galaxy images. However, this is not as accurate as other deblender outputs given that DeepDISC does not calculate pixel-level weighted segmentation maps.

\begin{figure*}[ht!]
    \centering
    \includegraphics[width=1\linewidth]{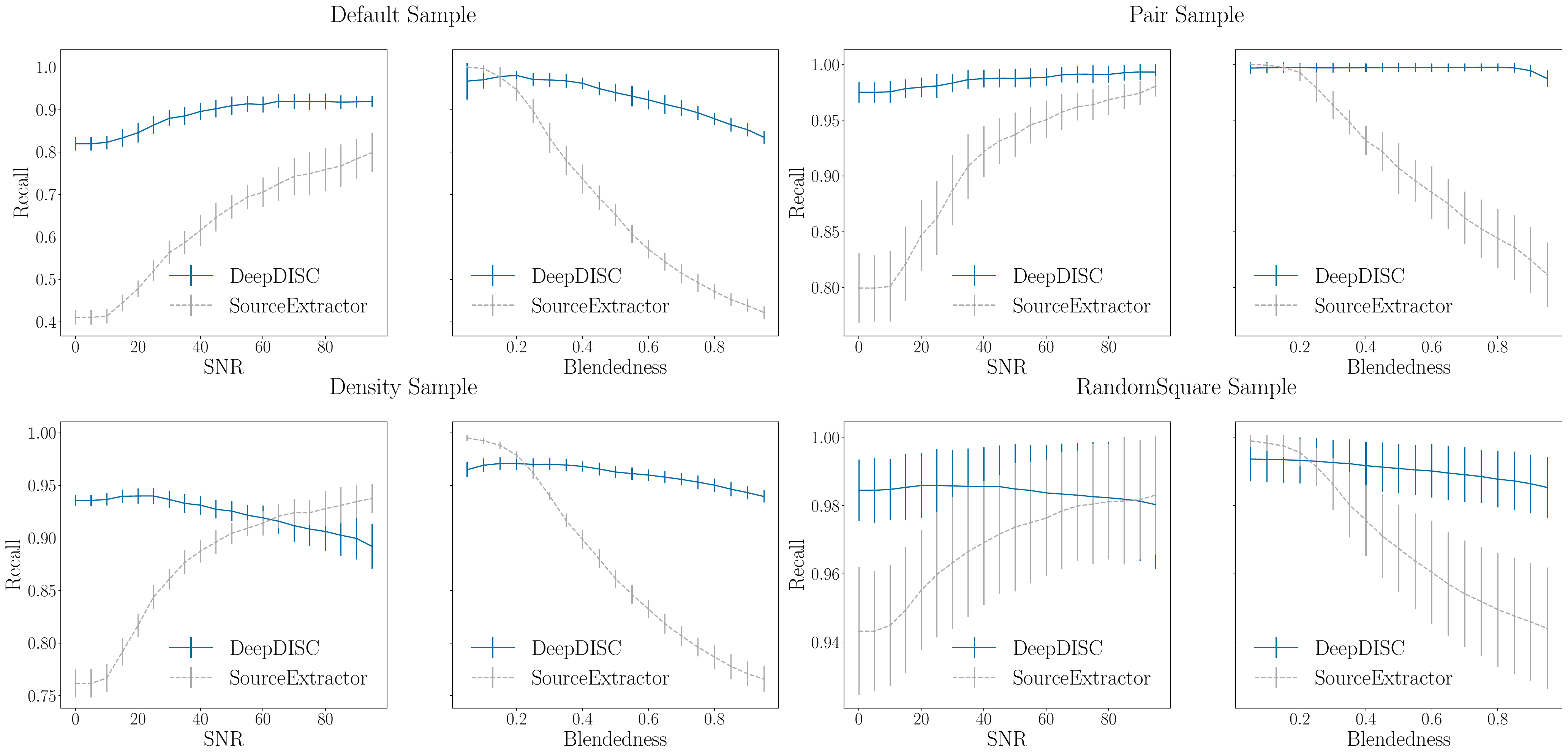}
    \caption{Comparison of Mean Recall Performance of DeepDISC (solid blue) and SourceExtractor (dashed grey) with varying image parameters of Signal-to-Noise Ratio (SNR) and Blendedness. Errors for each data-point were calculated using the bootstrap method. (Top Left) Recall performance with Default Sampled test images. (Top Right) Recall performance with Pair Sampled test images. (Bottom Left) Recall performance with Density Sampled test images. (Bottom Right) Recall performance with RandomSquare Sampled test images.}
    \label{fig:rec}
\end{figure*}
 
\subsection{SCARLET}\label{sec:scarlet}

SCARLET \citep{scarlet} is a deblending software used by many current and upcoming surveys. 
SCARLET employs Non-negative Matrix Factorization to deblend detected sources. It cannot detect sources in an image itself, and relies on other algorithms to provide a catalog of coordinates of potential sources as input. For LSST, an internal source detection pipeline provides SCARLET with coordinates to perform deblending. For this testing platform, the coordinates generated from GalSim are provided directly to SCARLET along with the blended image batch. Furthermore, this implementation of SCARLET is intended to reconstruct individual galaxies from a blend, thus not returning segmentation maps. The segmentation maps for SCARLET are made from its reconstructed output, using the same internal BTK thresholding method that identifies the ground truth segmentation from the simulated data sets.

The SCARLET implementation used here corresponds to the version described in \citet{scarlet}. A subsequent revision, \texttt{Scarlet-lite}\footnote{\url{https://github.com/lsst/scarlet_lite}}, introduces a more computationally efficient parameterization targeted at LSST-scale processing and has been adopted into the LSST Science Pipelines \citep{LSSTDP1}. In future benchmarks, implementing \texttt{Scarlet-lite} in combination with a source detection algorithm will allow for an emulation of the full LSST detection and deblending pipeline \citep{lsst_science_pipelines}.

\subsection{SourceExtractor}\label{sec:sextractor}

SourceExtractor \citep{sextractor} is a widely used source detection and feature extraction software. It has been used by several projects including the Dark Energy Spectroscopic Instrument Legacy Survey (DESI) \citep{desi_legacy}, the Pan-STARRS1 Survey \citep{pan_starrs}, and the Great Observatories Origins Deep Survey (GOODS) \citep{goods_survey}. 
It uses background estimation and thresholding to separate and ``extract" sources from images. It estimates the background by clipping the local background values until its distribution converges to around $\pm3\sigma$ of its median value. It then creates a background map using a mesh grid to spatially estimate the background using a median filter. SourceExtractor deblends sources using re-thresholding techniques to exponentially separate different levels of intensity from the initial threshold and the peak value. These levels are placed into a decision tree where an algorithm branches off to distinct sources that are spatially within in the initial source. A branch is made only if the integrated intensity at each level is greater than a specific fraction of the total intensity of the source.
% This is based on if the integrated intensity at each level is greater than a specified fraction of the total intensity of the initial source and that these conditions can be met at another branch at the same intensity level. 
This method is optimal for deblending single-band images where each source has the same intensity in that same passband. 

BTK uses SourceExtractor Python (SEP) \citep{SEP_2016} a Python wrapper API that calls the original source code written in C. We will refer to SEP as SourceExtractor throughout this work. 
% Multi-band efforts have been made within BTK, where the multi-band input images are fit individually and their identified sources are matched so that repeat detections can be removed and new sources preserved. 
For this benchmark we choose to use the single-band implementation with \textit{r}-band images. This is chosen to increase the functionality of SourceExtractor as background noise estimations for reconstructions can only be done with single-band images.
Single-band deep learning alternatives to SourceExtractor, such as the U-Net approach of \citet{Boucaud2020}, have demonstrated improved photometric recovery of high-redshift blended galaxies at LSST-like depths within a single bandpass. We also emphasize that this single-band implementation does not reflect how source detection is performed in the LSST Science Pipelines \citep{lsst_science_pipelines}, which employs multi-band coaddition and forced photometry across all six $ugrizy$ passbands. The SourceExtractor results here should therefore be interpreted as a classical algorithm baseline rather than a proxy for the operational LSST detection pipeline.

% The performance SExtractor for single-band images compared to multi-band images 

% \section{Results}\label{sec:results}

% The results are split into their respective metric sections where direct comparisons between the deblenders can be made. As each deblender has different capabilities and outputs, comparisons between each deblender and metric is not possible. Therefore, DeepDISC and SourceExtractor will be evaluated for detection performance and SourceExtractor and SCARLET will be evaluated for reconstruction. All deblenders will be compared and evaluated for segmentation. Furthermore, the performance section will provide information on the computational efficiency and timing of each deblender when given the same task.

% \begin{figure*}[ht!]
%     \centering
%     \includegraphics[width=1\linewidth]{figs/rec_plots_final1.pdf}
%     \caption{Comparison of Mean Recall Performance of DeepDISC (blue) and SourceExtractor (orange) with varying image parameters of Signal-to-Noise Ratio (SNR) and Blendedness. Errors for each data-point were calculated using the bootstrap method. (Top Left) Recall performance with Default Sampled test images. (Top Right) Recall performance with Pair Sampled test images. (Bottom Left) Recall performance with Density Sampled test images. (Bottom Right) Recall performance with RandomSquare Sampled test images.}
%     \label{fig:rec}
% \end{figure*}

\section{Results}\label{sec:results}

The results are split into their respective metric sections where direct comparisons between the deblenders can be made. As each deblender has different capabilities and outputs, comparisons between each deblender and metric is not possible. Therefore, DeepDISC and SourceExtractor are evaluated for detection performance and SourceExtractor and SCARLET are evaluated for reconstruction. All deblenders will be compared and evaluated for segmentation. Furthermore, Section \ref{sec:performance} provides information on the computational efficiency and timing of each deblender when given the same task.

\subsection{Detection}\label{sec:det_results}

Detection performance for SourceExtractor and DeepDISC is measured with Recall and Precision over bins of different Signal-to-Noise (SNR) ratios for the entire test set. Where
relevant, detection performance is also evaluated as a function of blendedness, defined here as the mean fraction of flux within a galaxy's aperture contributed by neighboring sources \citep{hsc_pipeline}. A blendedness of zero indicates an isolated source, while a value approaching unity indicates that most detected flux originates from overlapping neighbors. SNR for Recall is measured using the ground truth individual images of galaxies and are directly matched to true matched predictions. For Precision only, the SNR is measured for each detection output by the deblender instead of the SNR of true source generated by BTK. This is done to measure the SNR of all predictions made instead of only true matches which come from the ground truth. Each data point is the averaged value from a batch of images. Bootstrap errors were computed using $N = 1000$ resamples and represent the $1\sigma$ (68\%) confidence interval on the mean metric value per bin, shown to indicate the variance expected for any given image or detection. 

On average, DeepDISC has higher detection recall than SourceExtractor for both Default and Pair sampled images. For the Density sampled images the performance of DeepDISC is lower than of SourceExtractor and begins to decrease as the SNR increases (See Fig \ref{fig:rec}).  In cases of low SNR, DeepDISC has higher recall for the Default and Pair sampled images than SourceExtractor. As the SNR increases, SourceExtractor improves its performance and closely matches DeepDISC at SNR $\sim 90$. For images with many sources, supplied by the RandomSquare dataset, it is found that DeepDISC's recall performance is static with respect to SNR and blendedness. However, the recall values are higher than the previous data sets and is higher than SourceExtractor at low SNR and high blendedness, showing more robust detection performance with highly noised and dimmer sources. 

\begin{figure*}[ht!]
    \centering
    \includegraphics[width=1\linewidth]{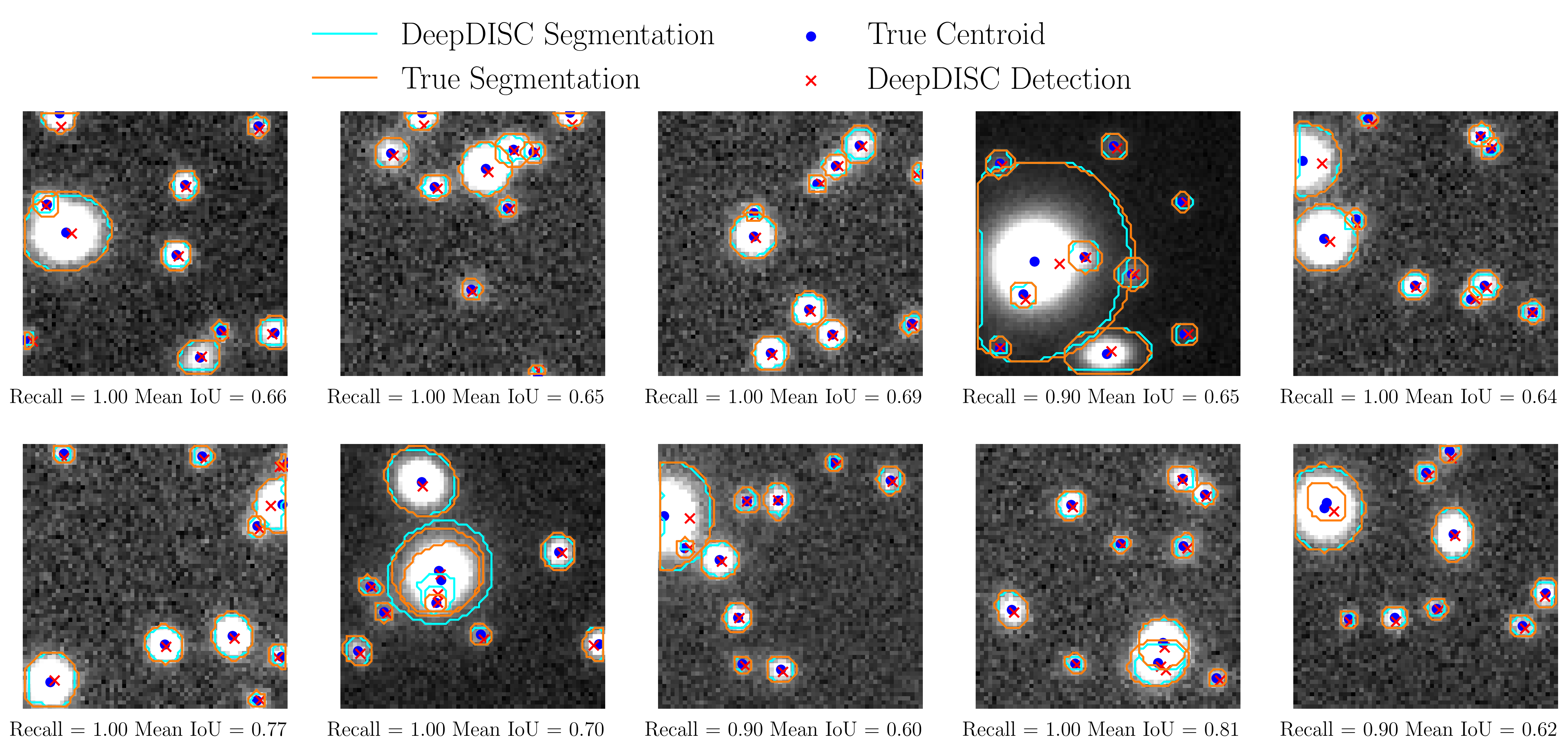}
    \caption{Random Density Sampling draws from DeepDISC's output showing Detection and Segmentation Performance compared with the ground truth. The teal curves show DeepDISC's predicted segmentation map and the orange curves show the true segmentation of each source. The blue points show the location of the true source centroid and the red ``X'' show DeepDISC's predicted centroid. The average Recall and IoU are listed below each image.}
    \label{fig:deepdisc_demo}
\end{figure*}

DeepDISC's performance could be explained by how it identifies galaxy centroids by using the center of a bounding box that surrounds the detection. For density sampled images, some sources are on the edge of images (See Fig \ref{fig:deepdisc_demo}) which skews the center of the bounding box and does not match the true centroid. This discrepancy creates many non-detections as the distance from the detected galactic center and the true center are outside of the matcher threshold. Further testing is required to rule out if DeepDISC's centroid detections are what is causing high residuals in crowded fields. For highly blended scenes, DeepDISC has shown to be more reliable in making accurate detections compared to SourceExtractor, however for scenes with high SNR SourceExtractor and DeepDISC's detection performance is comparable. 

\begin{figure*}
    \centering
    \includegraphics[width=1\linewidth]{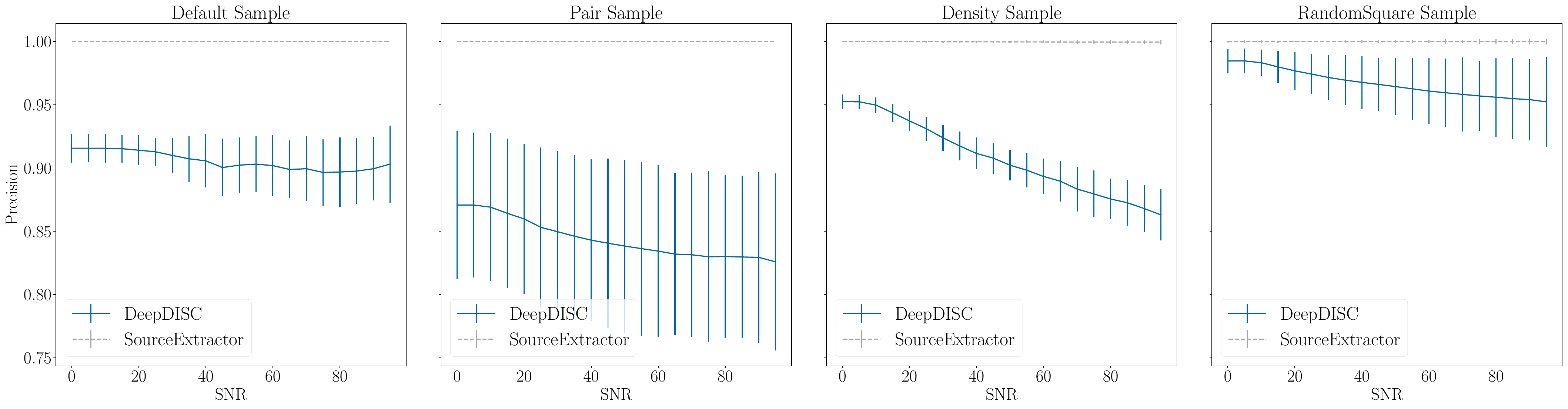}
    \caption{Precision performance of DeepDISC (solid blue) and SourceExtractor (dashed grey) with varying SNR values. Errors for each data-point were calculated using the bootstrap method. (Left) Precision performance for Default Sampled test images. (Middle-left) Precision performance for Pair Sampled test images. (Middle-right) Precision performance for Density Sampled test images. (Right) Precision performance for RandomSquare Sample test images.}
    \label{fig:prec}
\end{figure*}

For Precision performance, SourceExtractor has a close to constant value of 1 for each sample given (See Fig \ref{fig:prec}). SourceExtractor is a peak finding algorithm and can place centroids accurately for the sources that it does detect. As mentioned before, DeepDISC does not find peaks in the sources it detects, thus for the sources it does detect it may place the centers of sources offset from the true centroid used in matching. For Default Sampled images, DeepDISC varies its precision from around 0.92 to 0.9 and does not seem to have a trend with increasing SNR. For Pair Sampled images, DeepDISC has a downward trend with increasing SNR starting at around 0.88 and decreasing to 0.84. For the Density Sample, DeepDISC has high performance at low SNR at around 0.95, but decreases as the SNR increases ending at a value of 0.875 for 100 SNR. For the RandomSquare Sample, DeepDISC's performance is shown to be the highest out of the other test sets, however its precision still deceases with the increase of SNR.
The decreasing trends for Pair and Density Sampled images may indicate that DeepDISC's method for identifying the centroid of a source is less accurate as the sources become brighter. Larger sources being detected by DeepDISC have a higher chance of having an offset detected centroid that is larger than the true detection threshold. SourceExtractor's precision is consistently high due to its use of peak finding methods for deblending. Thus, for the sources that SourceExtractor does detect, its centroid is most likely going to be found. DeepDISC's lower precision values may be based on the peak finding limitation discussed earlier. It is recommended that DeepDISC's peak finding technique should be changed to be more robust, or to at least mark sources on the edges of images as non-detections.

% Based on SourceExtractor's peak finding detection method, its precision is high for the entire SNR range whereas DeepDISC has trouble with finding the true peak of the sources but had its highest performance with the largest resolution images and the most galaxies.

\subsection{Segmentation}\label{sec:seg_results}

\begin{figure}
    \centering
    \includegraphics[width=1\linewidth]{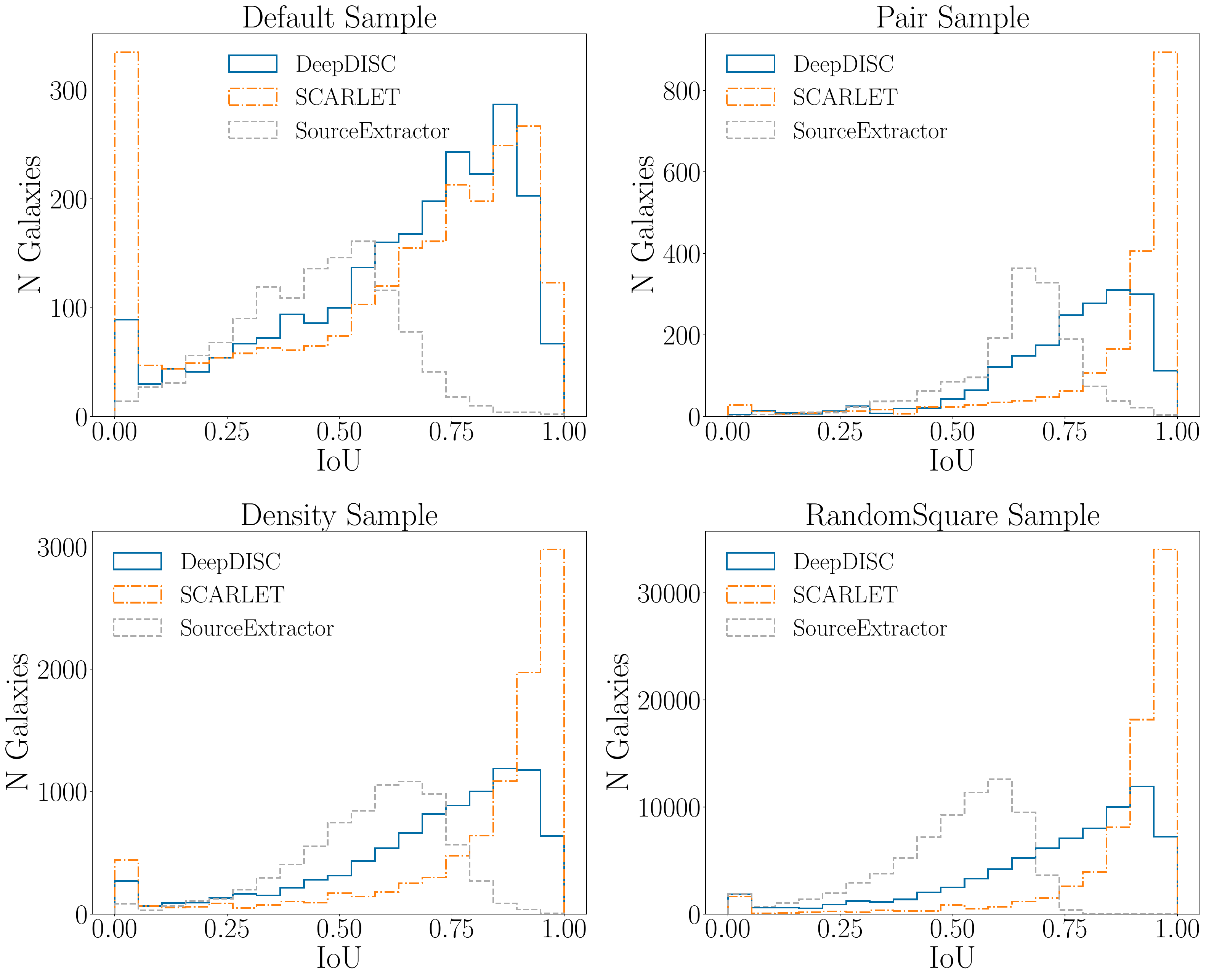}
    \caption{Distributions of Intersection-over-Union (IoU) values calculated from segmentation maps for each deblender: DeepDISC (solid blue), SCARLET (dash-dotted orange), and SourceExtractor (dashed grey). %(Left) Distribution of IoU values for Default Sampled test images. (Middle) Distribution of IoU values for Pair Sampled test images. (Right) Distribution of IoU values for Density Sampled test images.
    }
    \label{fig:iou}
\end{figure}

\begin{figure}[ht!]
    \centering
    \includegraphics[width=1\linewidth]{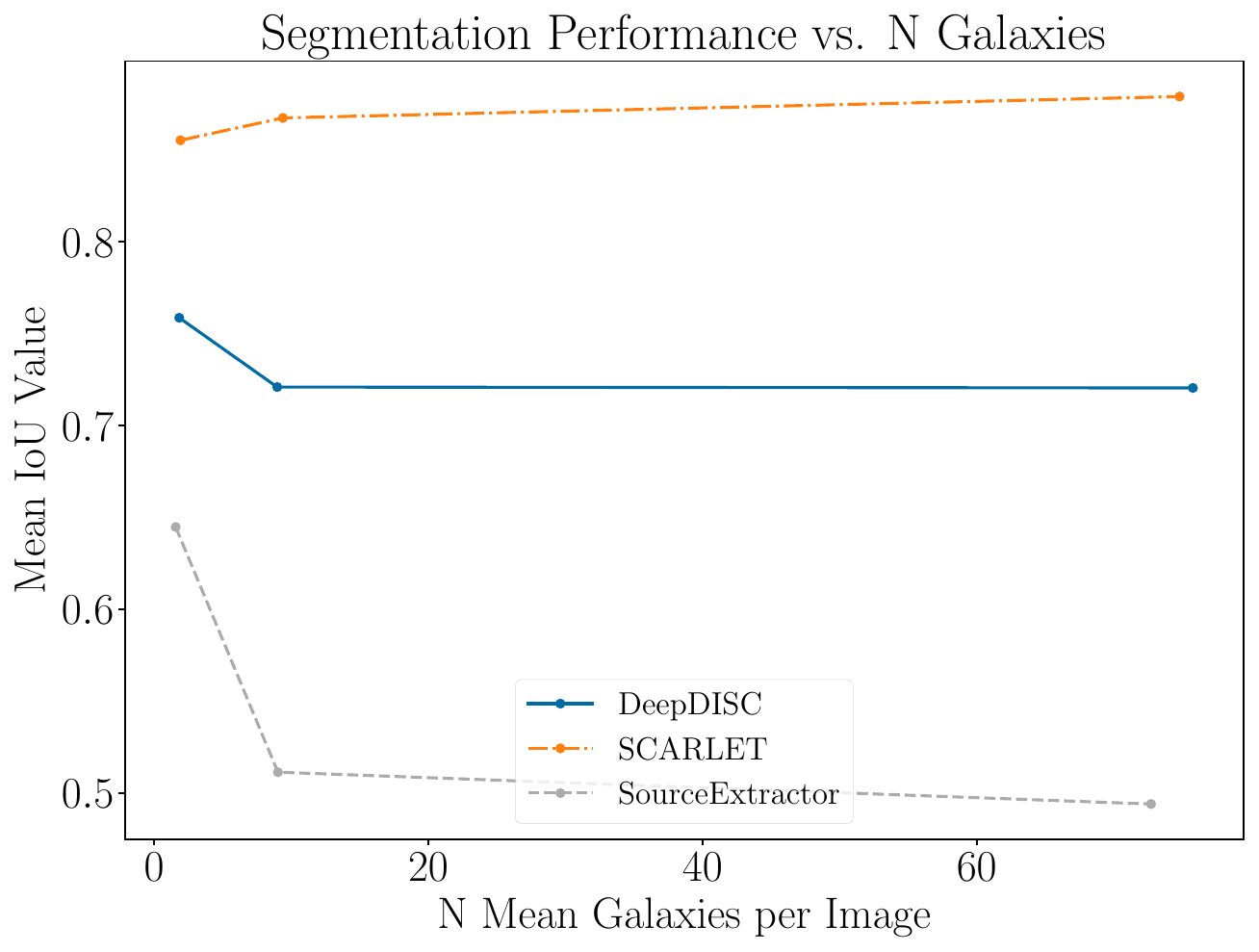}
    \caption{
    Segmentation performance as a function of mean galaxy density per image of DeepDISC (solid blue line), SCARLET (orange dash-dotted line), and SourceExtractor (grey dashed line). IoU values are drawn from the Pair, Density, and RandomSquare test datasets.}
    \label{fig:seg_perf}
\end{figure}

Segmentation performance between all deblenders is measured using the IoU of each predicted mapping. The distribution of IoU scores shows the overall performance of each deblender based on the sampling data used. When tested on the Density Sampled images, SCARLET and DeepDISC have many segmentation maps that have high IoU values ($0.8 < \text{IoU} < 1.0$) (See Fig \ref{fig:iou}). In particular, SCARLET has its peak close to an ideal IoU value of 1 with almost 4000 segmentation maps close to that value out of 9395 segmentation maps total. This high performance can be attributed to SCARLET's ability to reconstruct images of single galaxies from a blended scene, providing accurate segmentation. The median IoU value for DeepDISC in this distribution is 0.78 whereas SCARLET has a median value of 0.98 and SourceExtractor has a median IoU value of 0.54. For SourceExtractor's performance with RandomSquare Sampled images, none of segmentation maps it produced had a higher IoU value than 0.8. For the Pair Sampled test images, DeepDISC and SCARLET have similar performance to the Density Sampled images, whereas SourceExtractor has increased its performance (See Fig \ref{fig:iou}). Their medians are DeepDISC $=0.8$, SCARLET $=0.94$, and SourceExtractor $=0.66$. The Density Sample segmentation performance is comparable with the RandomSquare Sample (See Fig \ref{fig:seg_perf}). For the Default Sample distribution, SCARLET has many values with IoU's close to zero. This is due to the method used to extract segmentation maps from its output. Some of the galaxies that SCARLET generates are too faint or too large for a segmentation map to be correctly identified or drawn to be much larger than the source. This does not impact the reconstruction performance (See Section \ref{sec:recon_results}) however. For Segmentation performance DeepDISC is most reliable in the highly blended regime whereas SCARLET has the highest performance for all other regimes.
% \texthl{This indicates for DeepDISC that it has more conservative segmentations in larger fields as it has IoU values very close to 1 for each sample.} 
It is also shown that SourceExtractor has an upper performance limit of around 80\% coverage as the number of galaxies increases towards LSST scales. 

It is shown that SCARLET's segmentation performance increases as more sources are added to a scene, whereas DeepDISC's and SourceExtractor's performance decreases. SCARLET has the largest number of high-IoU sources for each sample tested. However, this can be attributed to SCARLET's output producing reconstructed images instead of segmentation maps. This allows SCARLET to generate highly detailed segmentation maps that can closely resemble the ground truth. In galaxy fields, sources may be attenuating less pixels allowing for smaller less detailed true segmentation maps. This may allow for higher IoU values for SCARLET. Furthermore, this performance discrepancy between the Pair Sample and higher density data like RandomSquare stems from a difference in source brightness and size. If the center source is bright and the offset source is considerably dimmer, deblenders like SourceExtractor may not recognize the offset source or identify it as a piece of the center source, returning only a single source 
% (i.e. unrecognized blend without blending) 
(see Pair Sample Recall Plot in Figure \ref{fig:rec}). This makes the IoU and Recall values at low SNR to be lower than expected and the same values to be larger at high SNR. This may explain SourceExtractor's segmentation threshold issue as more sources are included.
% as its IoU values have an 80\% threshold. 
Segmentations from SourceExtractor are either partially covering the true source, which is unlikely due to the nature of its deblending process, or that the map is enveloping other sources making it larger than the truth. This may be attributed to a constant intensity threshold for SourceExtractor that limits its segmentations. Specific tuning of its parameters may be required to accurately identify each source.

%This issue is planned to be rectified in following works on this subject. 
% \texthl{SCARLET had the largest number of high-IoU sources for each sample tested. However, this can be attributed to SCARLET's output being relegated to producing reconstructed images instead of segmentation maps. This allows SCARLET to generate highly detailed segmentation maps that can closely resemble the ground truth.}

% Fix default sampling problem with Scarlet then write here.

\subsection{Reconstruction}\label{sec:recon_results}

\begin{figure}
    \centering
    \includegraphics[width=1\linewidth]{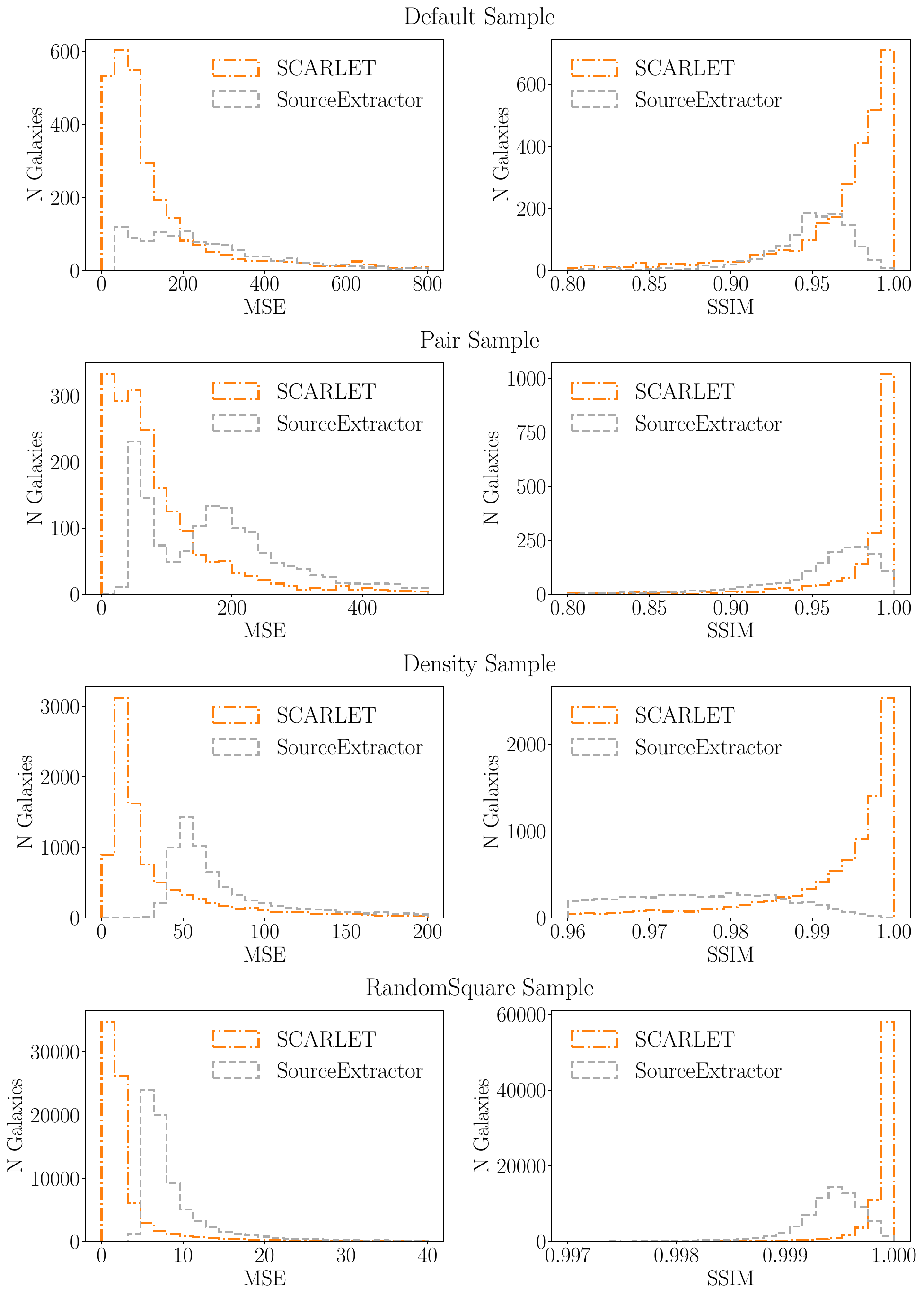}
    \caption{Distributions of the Mean Squared Error (MSE) and Structural Similarity Index (SSIM) of the reconstructed images output from SCARLET (dash-dotted orange) and SourceExtractor (dashed grey). %(Top) MSE and SSIM distribution for Default Sampled test images. (Middle) MSE and SSIM distribution for Pair Sampled test images. (Bottom) MSE and SSIM distribution for Density Sampled test images.
    }
    \label{fig:recon}
\end{figure}

Reconstruction performance is measured by pixel-level error (MSE) and quantitative shape and brightness measurements (SSIM). The distribution of the MSE and the SSIM for each galaxy reconstruction shows the performance of the deblenders given specific datasets (See Fig \ref{fig:recon}). For Default Sampled images, SCARLET had high performance with the peak of its distribution less than 100 MSE. SourceExtractor had a distribution with a median MSE value at around 230, compared to SCARLET having a median of around 84. For SSIM performance, both deblenders have high performance with most reconstructions within 0.8. 
% SourceExtractor still has a broader distribution than SCARLET \texthl{with a Median Absolute Deviation}, but has most of its values close to 1. 
Their respective medians for the SSIM distribution are SCARLET $=0.97$ and SourceExtractor $=0.95$. For Pair Sampled images, there is a large gap in the MSE distribution for SourceExtractor (See Fig \ref{fig:recon}). SCARLET has a median MSE of 64 where SourceExtractor has a median MSE value of 179. For SSIM, SCARLET has a median value of 0.99 and SourceExtractor has a median of 0.96. 
For density sampled images, both deblenders have significantly improved MSE and SSIM values when compared to the other sampled image sets. 
% Both deblenders had their lowest median MSE values with the density sampled images.
SCARLET has a median MSE value of 20 and SourceExtractor has a median MSE of 65 for density sampled images. For SSIM when tested the density sample, both deblenders have the closest median SSIM values with SCARLET having a median SSIM of 0.994 and SourceExtractor having a median of 0.967. For the RandomSquare Sample, SourceExtractor's MSE distribution is the least broad out of the other test sets with a median value of $\sim7$ and a Median Absolute Deviation value of 1.45, compared to the Default Sample at 117.1. However, SCARLET's MSE median value decreased compared to the density sample $=1.7$. The SSIM values for the RandomSquare Sample are comparable to the Density Sample with median values for SCARLET $=0.999$ and SourceExtractor $=0.999$. Statistical values from these distributions are shown in \autoref{tab:recon_stat_vals}.

SCARLET has high performance in source reconstruction with many sources with low MSE and high SSIM values. SCARLET performs best with the Density and RandomSquare datasets. As mentioned earlier, SCARLET outputs reconstructed galaxies far dimmer than the truth when given Default Sampled images. For other reconstructed sources, their profiles are sometimes brighter and larger than usual, indicating that there is an attempt for the total flux to be conserved. This made generating their segmentation maps difficult. This limitation may signify that sources reconstructed by SCARLET in heavily blended scenes may not reflect the true brightness or extent of the source. This may be influenced by passing SCARLET true detection centroids of highly blended sources. These sampled ground truth centroid locations may be closer than what SCARLET is built to expect given that it is intended to work in tandem with a source detection pipeline which may not be as precise. This results in SCARLET reconstructions that blend multiple bright sources together and does not fully reconstruct dimmer sources in the original blend. For SourceExtractor, its reconstructions are done with a ``cookie-cutter" method, where it is cut from the segmentation maps directly \citep{sextractor}. SourceExtractor's reconstruction performance is highly tied to its segmentation performance, which has been found to create many unrecognized blends for highly blended sources. 
The perceived performance increase for both deblenders when given the RandomSquare sample can be attributed to the method of calculating SSIM and MSE. Both of these metrics are calculated using the ground truth images of individual galaxies and the deblender reconstructions. The RandomSquare dataset covers a larger on-sky area, which results in reconstructed images with more background than the first 3 sample datasets. This increased background area may increase the similarity between the ground truth and the reconstruction.

\begin{table*}[ht!]
    \centering
    \begin{tabular}{m{2cm}|ccc|ccc||ccc|ccc|}
        & \multicolumn{6}{c||}{Mean Squared Error}&\multicolumn{6}{c|}{Structural Similarity Index Metric}\\\cline{2-13}
          & \multicolumn{3}{c|}{SCARLET}& \multicolumn{3}{c||}{SourceExtractor}& \multicolumn{3}{c|}{SCARLET}& \multicolumn{3}{c|}{SourceExtractor} \\
         Sample & Median & MAD & N & Median & MAD & N & Median & MAD & N & Median & MAD & N\\\hline
         Default & 84.35 & 52.02 & 2837 & 230.40 & 117.10 & 1136&0.978 & 0.015 & 2834&0.952 & 0.014 & 1220\\
         Pair& 64.18 & 41.43 & 1900 & 178.08 & 88.69 & 1480&0.992& 0.0066 & 1850&0.965 & 0.017& 1552\\
         Density& 20.02 & 10.93 & 9325 & 65.19 & 17.82 & 6714&0.994 & 0.0043 & 8808&0.967 & 0.013& 4867\\
         RandomSquare& 1.74 & 0.66 & 77690 & 7.18 & 1.45 & 72977&0.99948 & 3.95e-05 & 78259&0.99945 & 0.00017 & 74047\\\hline
    \end{tabular}
    \caption{Statistical Values from Reconstruction Metric Distributions. Median, Median Absolute Deviation (MAD), and total number of galaxies (N) are shown for both the Mean Squared Error (MSE) and Structural Similarity Index Metric (SSIM) Distributions.}
    \label{tab:recon_stat_vals}
\end{table*}

% It is shown that SCARLET's reconstruction performance increases as the number of galaxies in each image increases. At larger scales with RandomSquare sampled images, SCARLET and SourceExtractor have comparable performance, where the errors for both deblenders are negligible.

\subsection{Computation Time}\label{sec:performance}

Compute time\footnote{Tested on a NCSA Delta A100x4 GPU single CPU compute node with 16 GB memory.} for a single batch of 10 images show the test wall time for deblending the datasets used in the benchmark\footnote{RandomSquare Sample is not listed due to extended computation times.} (See Table \ref{tab:timing}). For each sampling image test set, the deblenders process a set of Default Sampled images the fastest with Pair Sampled and Density Sampled images being overall slower. The fastest deblender is SourceExtractor which uses about half of DeepDISC's computation time on average. The slowest deblender is SCARLET which is found to be $\sim7$ times the computation time of DeepDISC. It should be noted that \texttt{Scarlet-lite}, a more computationally efficient version of SCARLET and used within the LSST Pipeline, is not benchmarked here. 
% Additionally, DeepDISC caches some initialization files \texthl{which makes the} initialization of the first deblend job take longer. The initialization time is removed from this analysis.

\renewcommand{\arraystretch}{1.5}
\begin{table}[ht!]

    \centering
    \begin{tabular}{m{2.2cm}|c|c|c}
    % \hline
    % \hline
    Deblender & Default (sec) & Pair (sec) & Density (sec)\\ \hline \hline
    DeepDISC &  1.193123&1.840722&2.643613\\
    SCARLET& 8.394066&12.351854&17.403525\\
    SourceExtractor & 0.507302&0.636077&1.268854\\ \hline
    \end{tabular}
    
    \caption{Computation times for deblending a single batch of 10 images from each data sample.}
    \label{tab:timing}
\end{table}

\section{Discussion}
\label{sec:discussion}

\setlength{\tabcolsep}{12pt}
\renewcommand{\arraystretch}{2}

\begin{table*}[ht]
\centering
\caption{Summary of strengths and limitations by deblender.}
\begin{tabular}{p{3cm} p{4.2cm} p{4.2cm} p{3.4cm}}
\hline
\textbf{Deblender} & \textbf{Key Strengths} & \textbf{Primary Limitations} & \textbf{Best Use Case} \\
\hline
SourceExtractor & Fast; precise centroiding; mature and stable & Low recall for faint/blended sources; poor segmentation in dense scenes & Bright, sparse fields; centroid-critical tasks \\
SCARLET & Excellent segmentation and reconstruction; strong in dense fields & Requires external detection; computationally expensive & High-fidelity reconstruction in crowded regions \\
DeepDISC & High recall at low SNR; integrated detection and segmentation; scalable & Centroid bias in dense fields; training-data dependence & Faint, blended source detection at LSST scale \\
\hline
\end{tabular}
\label{tab:deb_streng_lims}
\end{table*}

\subsection{Strengths and Limitations}

% \texthl{Refer to }\autoref{tab:deb_streng_lims}.

Each deblender has their own strengths and limitations. Some recommendations can be made about ideal use cases based on the benchmarking results (See \autoref{tab:deb_streng_lims}). 

DeepDISC is found to have high recall with detections and robust segmentation performance throughout all testing cases. Its limitation on identifying centroid locations is something that can be improved with future iterations and is not a fundamental flaw with the model. However, DeepDISC's dependence on a trained machine learning model does have an impact on performance, as with any machine learning based process. From this benchmark, the choice to use sparse and randomized data (e.g. Density and RandomSquare) increased its performance. DeepDISC's computational performance allows it to be scaled up to process large survey data. Furthermore, DeepDISC is shown to have consistent Detection performance across the SNR range, allowing it to pick up on fainter sources.

SCARLET's performance is high for segmentation and reconstruction, however it is computationally intensive. Its segmentation performance is consistent with high IoU values throughout. Likewise, its reconstruction performance is also consistently high for each dataset sample. This allows SCARLET to be reliable in crowded scenes with multiple sources. However, its computational cost rises significantly with an increase in sources compared to other deblenders (See \autoref{tab:timing}). This limitation is intrinsic to SCARLET's statistics based deblending methods. As mentioned earlier, using specific iterations of SCARLET built for large surveys (i.e. \texttt{Scarlet-lite}) may mitigate these computational costs.

% improvements to SCARLET have been made potentially mitigating the computational cost.

SourceExtractor is a strong peak-finding method that is computationally efficient but has consistently low performance in most regimes. Its performance for low SNR sources indicate a higher chance to miss dimmer sources when compared to DeepDISC. Its recall does increase as the SNR increases, with higher values similar to DeepDISC. Combining these results with its segmentation performance show that SourceExtractor is most likely enveloping multiple sources and identifying the peaks of the brighter source. This limitation may produce unrecognized blends if used unsupervised in a survey processing setting. This limitation may be mitigated by modifying its thresholding parameter for bright sources with extended profiles. This benchmark is not intended to identify the best parameters for each deblender. On small scales, SourceExtractor is ideal for fast computation compared to other deblenders that require external factors like training a model or a pre-existing detection catalog. It should be stated that using single-band observations may impact its performance in this benchmark, however all ground truth comparisons made with SourceExtractor use only single-band data.

% \subsection{\textbf{Limitations of Evaluation}}

% \texthl{Talk about caveats here.}

\subsection{Expanding BTK}\label{sec:expand_btk}

% \texthl{Talk about expanding BTK here.}
% \newpage
This project's results are limited by several constraints. These limitations in simulation quality and deblender implementation restrict our ability to fully characterize deblender performance in large surveys. Expanding BTK with more metrics and deblenders would allow for stronger conclusions to be made. 
% Expansions to BTK have begun with the inclusion of DeepDISC for this benchmark.
% The first extension of the BTK since its publication was the addition of DeepDISC for this project. 
Extending this framework to include additional deblenders (e.g., BLISS \citep{bliss,Mendoza2026}, MADNESS \citep{madness}, SCARLET2 \cite{Ward2025}), more complex blending scenarios, and science-driven performance metrics would allow for greater recommendations to be made. Furthermore, the inclusion of an LSST-like detection pipeline alongside SCARLET would provide better insight into the performance of the LSST detection and deblending process.

% Expanding the evaluation metrics and image simulation processes would also provide greater insights to the benchmark. Including metrics based on photometric redshifts and other analysis methods using galaxy properties would allow the benchmark to show potential impacts on cosmological analyses. Furthermore, including detailed shape analysis methods based on segmentation shapes could provide direct evaluations on lensing shape measurements. This requires increasing the detail of the ground truth data utilized and incorporating cosmological analysis pipelines into the BTK. 

As discussed earlier, the ability to deblend galaxies is directly linked to the accuracy of several cosmological analyses. Understanding the impacts of deblending performance on downstream cosmological analyses would allow for direct comparisons to cosmological observables. Including metrics based on redshift and shape measurements would allow the benchmark to identify potential uncertainties when photometric redshift and weak lensing algorithms use deblended data. This requires increasing the detail of the ground truth data and incorporating cosmological analysis pipelines into the BTK. We make suggestions below for a DESC data challenge that will incorporate stronger galaxy attributes into ground truth simulations.

% Increasing the amount of ground truth information utilized in the benchmark will also allow for greater understanding of how potential downstream analyses would be impacted. 

% \subsection{Implications for LSST Science Pipelines}

% The prevalence of unrecognized blends in SourceExtractor and, to a lesser extent, DeepDISC segmentation outputs underscores a critical risk for cosmological and extragalactic analyses \citep{Sheldon2020,Prat2023}. 
% %LSST weak-lensing and photometric-redshift analyses.
% Unrecognized blends can bias shape measurements and flux estimates in ways that are difficult to diagnose post hoc, particularly when blends are treated as single objects throughout the pipeline \citep{Mandelbaum2018b,Nourbakhsh2022,liang_adari_unrec_blends}. While SCARLET mitigates this risk through high-fidelity reconstruction, its dependence on an upstream detection catalog means that detection failures propagate directly into reconstruction performance.

% These results naturally motivate hybrid pipeline strategies, for example using DeepDISC for initial detection and blend identification, followed by SCARLET for reconstruction in dense regions. Such approaches may offer a practical balance between recall, reconstruction fidelity, and computational cost, and they align well with LSST's modular pipeline architecture.

\subsection{Deblending Data Challenge}\label{sec:data_challenge}

This work is intended as a foundation for a broader DESC-led deblending data challenge. 
% By leveraging BTK to generate reproducible, LSST-like blended scenes with known ground truth, we demonstrate that end-to-end, regime-aware benchmarking is both feasible and scientifically informative. Extending this framework to include additional deblenders (e.g., BLISS \citep{bliss,Mendoza2026}, MADNESS \citep{madness}, Scarlet2 \cite{Sampson2024}), more complex blending scenarios, and science-driven performance metrics would enable the community to systematically evaluate deblending strategies under conditions directly relevant to LSST cosmology analyses. 
% Improving blending simulations would allow for greater analysis on deblending
Expanding blending simulations would enable more comprehensive characterization of deblender performance. Improvements to color accuracy, lensing effects, and observational phenomena would provide more realism to ground truth data. Current simulations, like GalSim and DC2, have specific limitations including drawing clusters with similar colors, and not fully emulating lensing effects. Likewise, PSF modeling and other instrumental effects are bulk modeled, which doesn't account for effects such as atmospheric turbulence. Improving the modeling of these effects in simulations would allow for a greater understanding of how deblenders work in a realistic survey setting.
% \texthl{Understanding the uncertainties that deblending creates in a large survey allows the community to identify }
% \texthl{}
% \texthl{Furthermore, the challenge should increase the number of unique blended scenes to match the number of potential observations from LSST. This increased depth would allow for the data to be incorporated into larger simulated survey datasets.}

With these simulation improvements as a foundation, a community challenge built on this framework would serve multiple purposes: establishing standardized evaluation datasets and metrics, clarifying trade-offs between competing approaches, and accelerating the development of hybrid or next-generation deblenders. In this sense, the present study should be viewed not as a final ranking of methods, but as a baseline benchmark and methodological template upon which a community-wide effort can build.

\section{Conclusion}
\label{sec:concl}

In summary, these findings have shown both specific areas for improvement and the overall uncertainties when using each deblender. Overall, DeepDISC had higher recall performance at low SNR when compared to SourceExtractor, however DeepDISC had trouble performing in dense galaxy fields. This is likely due to the bounding-box centroid issue in DeepDISC. Quantification of the centroid error in DeepDISC will be investigated with additional testing outside of the main BTK framework. Furthermore, DeepDISC was trained using Density and RandomSquare Sampled image sets, which are generated using the same method which creates the benchmark datasets, which may indicate a bias in performance. SCARLET had overall high segmentation and reconstruction performance, but had persistent issues with reconstructing highly blended sources. Discrepancies between the true source and specific reconstructed sources in terms of brightness are found when using SCARLET on highly blended images. This may be an impact of passing SCARLET true galaxy centroids which may place source centers too close for effective deblending. The performance of SourceExtractor for each metric was lower than both DeepDISC and SCARLET for many blending situations. SourceExtractor's methods for detection allow it to be highly precise in centroid finding, but not ideal for detecting all the sources in an image. Its segmentation performance was shown to be consistently low with many unrecognized blends and partial segmentation coverage. Its methods for reconstruction, which are mostly based on segmentation were also shown to be lower than SCARLET in both SSIM and MSE values. It should be noted that SourceExtractor is only deblending single r-band images, and is taken in account when evaluating its performance, however both DeepDISC and SCARLET both use multi-band images when deblending.

% \texthl{Due to the enclosed nature of the benchmark, certain decisions have been made to the benchmarking process that may impact specific deblender performance.}
% The low performance in detection, segmentation, and reconstruction on highly blended sources indicates large uncertainties for each deblender. 

% \subsection{Future Work}

Expanding this work after the conclusion of this initial benchmarking report is a priority for the authors. We plan to expand the testing parameters by including more complex blended scenes in new simulated sample image test sets. Furthermore, the inclusion of new deblenders for benchmarking like the Bayesian Light Source Separator \citep[BLISS;][]{bliss} and the Maximum A posteriori with Deep NEural networks for Source Separation \citep[MADNESS;][]{madness} is also a priority. Working with deblender developers to integrate their algorithms into the BTK allows them to validate their own work and to compare them to other algorithms. Since SCARLET is integrated into the LSST pipeline, emulating LSST's detection algorithm so it can work alongside SCARLET in the BTK can provide the detection performance of the LSST analysis pipeline. A publication with this expanded work is expected in the future.

\section*{Acknowledgments}

A.B. led the software development, data curation, analysis, visualizations, and paper writing. G.M. advised A.B. on best practices for deblender integration, provided expertise, and contributed to writing the paper. X.L. advised A.B., aided in conceptualizations, provided visualizations, and contributed in writing the paper. 

We thank Dr. I. Mendoza for helpful discussion and calculation support. We thank Prof. Y. Wang for helpful discussion. We thank Drs. V. Kindratenko, S. Luo, and B. Bode at the National Center for Supercomputing Applications (NCSA) for helpful discussion and assistance with the GPU cluster used in this work. 

% DESC acknowledgement here
This paper has undergone internal review in the LSST Dark Energy Science Collaboration.
We would like to thank our reviewers Prakruth Adari and Cyrille Doux for their detailed and thorough comments.

A.B., G.M., and X.L. acknowledge support by Illinois Campus Research Board Award
RB25035, NSF grant AST-2308174, and NASA grants 80NSSC24K0219 and 80NSSC26K0333.
% for work using the HAL cluster:
This work utilizes resources supported by the National Science Foundation's Major Research Instrumentation program, grant \#1725729, as well as the University of Illinois at Urbana-Champaign. 
% NSF ACCESS blurb:
% ACCESS acknowledgement
This work used Delta and DeltaAI at NCSA through allocations PHY240290 and PHY250333 from the Advanced Cyberinfrastructure Coordination Ecosystem: Services \& Support (ACCESS) program, which is supported by U.S. National Science Foundation grants \#2138259, \#2138286, \#2138307, \#2137603, and \#2138296.
\input{standard_desc_ack}

We also acknowledge the use of \texttt{numpy} \citep{numpy}, \texttt{matplotlib} \citep{matplotlib}, and Astropy \citep{astropy_13}.

\section*{Data Availability}

The BlendingToolKit (BTK) has been modified from its original version with the addition of DeepDISC. This modified version can be found on GitHub as a fork of the original\footnote{\url{https://github.com/LSSTDESC/BlendingToolKit}} at the following link: \url{https://github.com/berres2002/BlendingToolKit}. 

\newpage
\bibliographystyle{aasjournal}

% You should give the same name for your .bbl as your main .tex
% since it is a requirement for posting on ArXiv.
\bibliography{main}

\newpage
\begin{appendix}
% \newpage
Here we show some further insights with the Segmentation and Reconstruction metrics and their correlation to the Signal-to-Noise ratio and Blendedness of the measured galaxies.
% \section{Appendix 1}
% \label{ap:ap}
\begin{figure*}[hb!]
    \centering
    \includegraphics[width=1\linewidth]{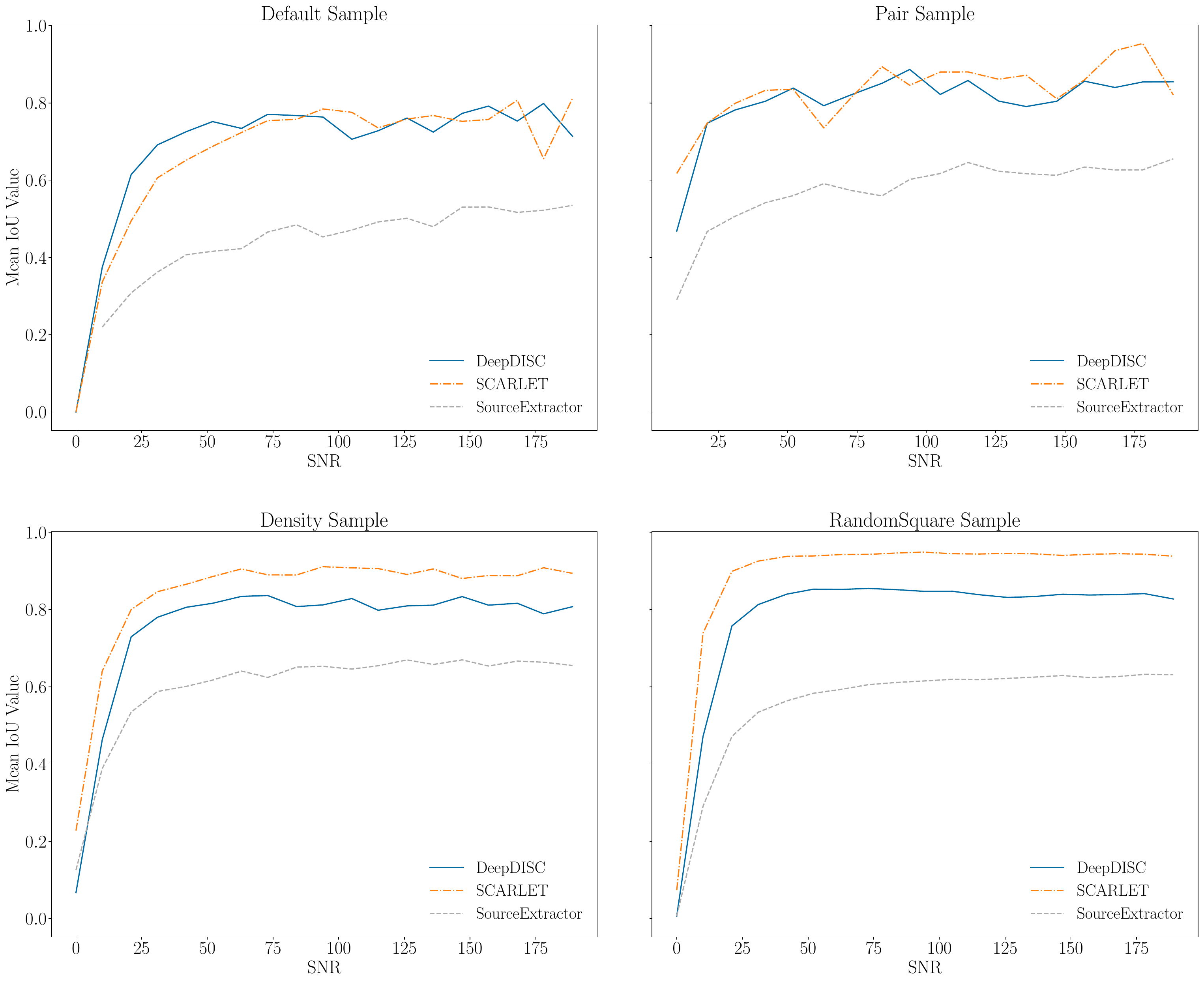}
    \caption{
    Mean Intersection over Union (IoU) Values of DeepDISC (blue solid line), SCARLET (orange dashed-dotted line), and SourceExtractor (grey dashed line) binned with respect to the Signal to Noise Ratio (SNR) of each IoU source.}
    \label{fig:iou_snr}
\end{figure*}

\end{appendix}

\begin{figure*}
    \centering
    \includegraphics[width=1\linewidth]{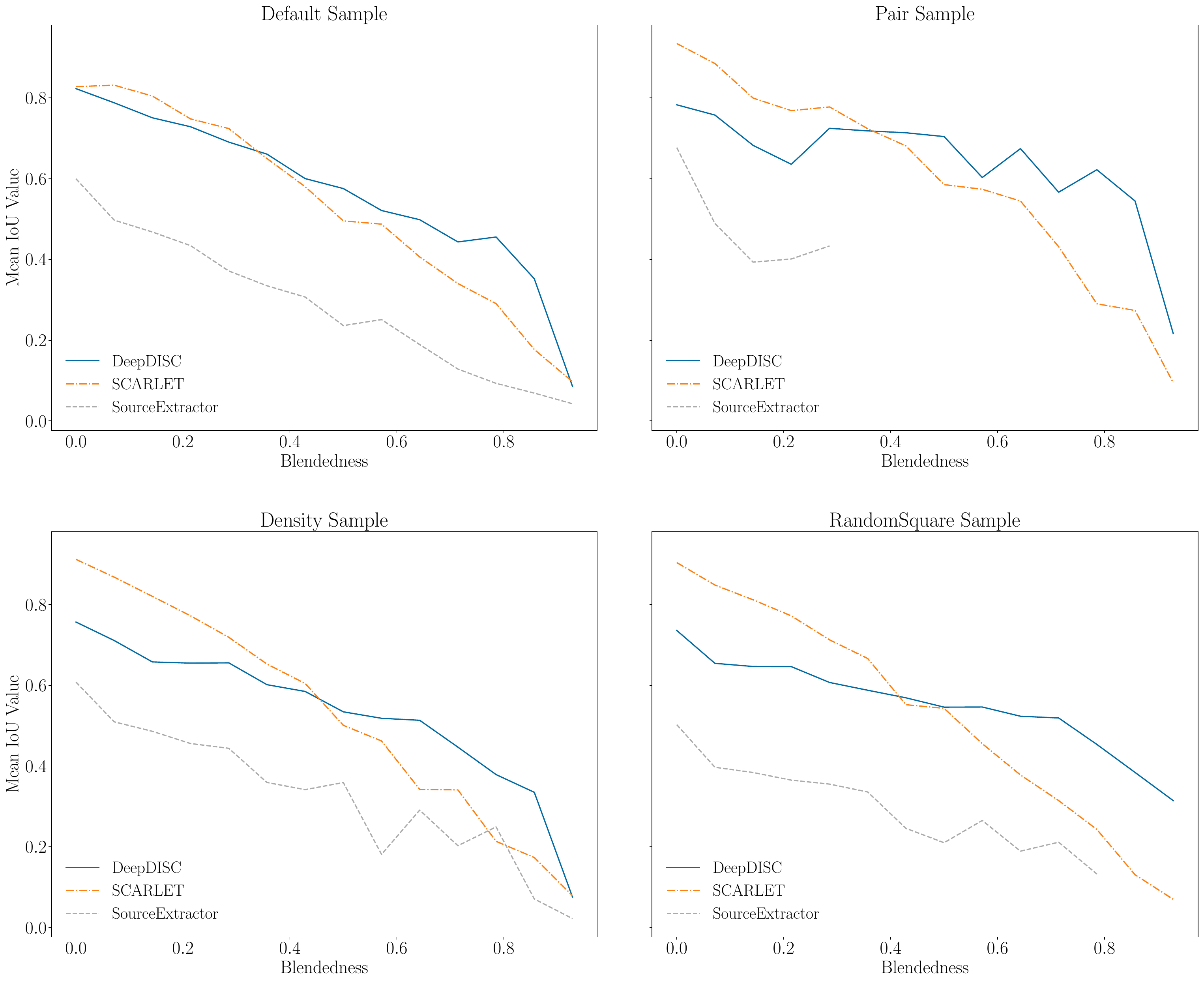}
    \caption{Mean Intersection over Union (IoU) Values of DeepDISC (blue solid line), SCARLET (orange dashed-dotted line), and SourceExtractor (grey dashed line) binned with respect to the Blendedness of each IoU source measured.} 
    \label{fig:iou_blend}
\end{figure*}

\begin{figure*}
    \centering
    \includegraphics[width=1\linewidth]{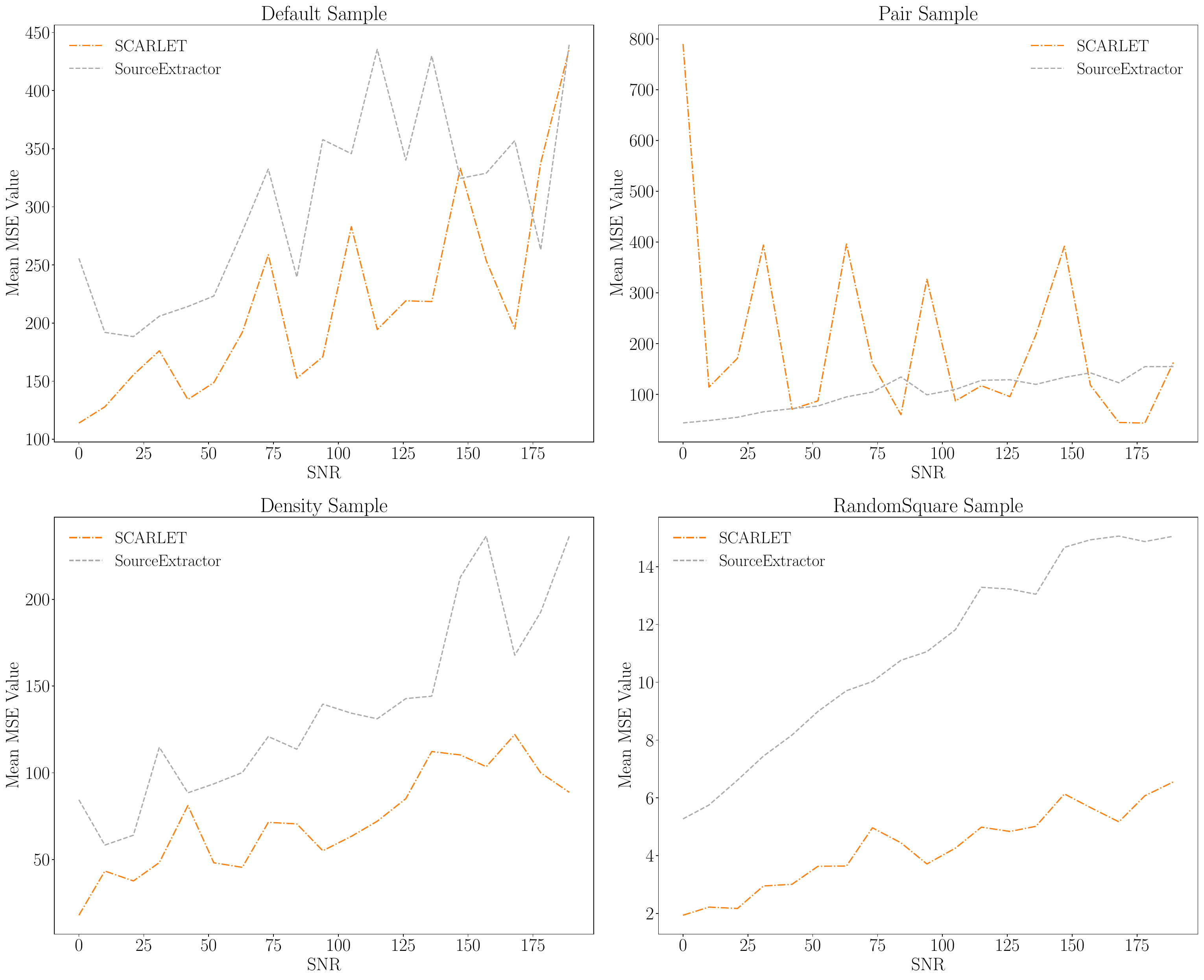}
    \caption{Mean Mean Squared Error (MSE) Values of SCARLET (orange dashed-dotted line), and SourceExtractor (grey dashed line) binned with respect to the Signal to Noise Ratio (SNR) of each source reconstructed.} 
    \label{fig:mse_snr}
\end{figure*}

\end{document}

%% file: standard_desc_ack.tex
The DESC acknowledges ongoing support from the Institut National de 
Physique Nucl\'eaire et de Physique des Particules in France; the 
Science \& Technology Facilities Council in the United Kingdom; and the
Department of Energy and the LSST Discovery Alliance
in the United States.  DESC uses resources of the IN2P3 
Computing Center (CC-IN2P3--Lyon/Villeurbanne - France) funded by the 
Centre National de la Recherche Scientifique; the National Energy 
Research Scientific Computing Center, a DOE Office of Science User 
Facility supported by the Office of Science of the U.S.\ Department of
Energy under Contract No.\ DE-AC02-05CH11231; STFC DiRAC HPC Facilities, 
funded by UK BEIS National E-infrastructure capital grants; and the UK 
particle physics grid, supported by the GridPP Collaboration.  This 
work was performed in part under DOE Contract DE-AC02-76SF00515.

%% file: main.bib
@ARTICLE{btk,
       author = {{Mendoza}, Ismael and {Torchylo}, Andrii and {Sainrat}, Thomas and {Guinot}, Axel and {Boucaud}, Alexandre and {Paillasa}, Maxime and {Avestruz}, Camille and {Adari}, Prakruth and {Aubourg}, Eric and {Biswas}, Biswajit and {Buchanan}, James and {Burchat}, Patricia and {Doux}, Cyrille and {Joseph}, Remy and {Kamath}, Sowmya and {Malz}, Alex I. and {Merz}, Grant and {Miyatake}, Hironao and {Roucelle}, C{\'e}cile and {Zhang}, Tianqing and {LSST Dark Energy Science Collaboration}},
        title = "{The Blending ToolKit: A simulation framework for evaluation of galaxy detection and deblending}",
      journal = {The Open Journal of Astrophysics},
         year = 2025,
        month = feb,
       volume = {8},
        pages = {E14},
          doi = {10.33232/001c.129699},
archivePrefix = {arXiv},
       eprint = {2409.06986},
 primaryClass = {astro-ph.IM},
       adsurl = {https://ui.adsabs.harvard.edu/abs/2025OJAp....8E..14M}
}

@ARTICLE{dawson_16,
       author = {{Dawson}, William A. and {Schneider}, Michael D. and {Tyson}, J. Anthony and {Jee}, M. James},
        title = "{The Ellipticity Distribution of Ambiguously Blended Objects}",
      journal = {\apj},
         year = 2016,
        month = jan,
       volume = {816},
       number = {1},
          eid = {11},
        pages = {11},
          doi = {10.3847/0004-637X/816/1/11},
archivePrefix = {arXiv},
       eprint = {1406.1506},
 primaryClass = {astro-ph.CO},
       adsurl = {https://ui.adsabs.harvard.edu/abs/2016ApJ...816...11D}
}

@ARTICLE{scarlet,
       author = {{Melchior}, P. and {Moolekamp}, F. and {Jerdee}, M. and {Armstrong}, R. and {Sun}, A. -L. and {Bosch}, J. and {Lupton}, R.},
        title = "{SCARLET: Source separation in multi-band images by Constrained Matrix Factorization}",
      journal = {Astronomy and Computing},
         year = 2018,
        month = jul,
       volume = {24},
          eid = {129},
        pages = {129},
          doi = {10.1016/j.ascom.2018.07.001},
archivePrefix = {arXiv},
       eprint = {1802.10157},
 primaryClass = {astro-ph.IM},
       adsurl = {https://ui.adsabs.harvard.edu/abs/2018A&C....24..129M}
}

@ARTICLE{sextractor,
       author = {{Bertin}, E. and {Arnouts}, S.},
        title = "{SExtractor: Software for source extraction.}",
      journal = {\aaps},
         year = 1996,
        month = jun,
       volume = {117},
        pages = {393-404},
          doi = {10.1051/aas:1996164},
       adsurl = {https://ui.adsabs.harvard.edu/abs/1996A&AS..117..393B}
}

@ARTICLE{galsim,
       author = {{Rowe}, B.~T.~P. and {Jarvis}, M. and {Mandelbaum}, R. and {Bernstein}, G.~M. and {Bosch}, J. and {Simet}, M. and {Meyers}, J.~E. and {Kacprzak}, T. and {Nakajima}, R. and {Zuntz}, J. and {Miyatake}, H. and {Dietrich}, J.~P. and {Armstrong}, R. and {Melchior}, P. and {Gill}, M.~S.~S.},
        title = "{GALSIM: The modular galaxy image simulation toolkit}",
      journal = {Astronomy and Computing},
         year = 2015,
        month = apr,
       volume = {10},
        pages = {121-150},
          doi = {10.1016/j.ascom.2015.02.002},
archivePrefix = {arXiv},
       eprint = {1407.7676},
 primaryClass = {astro-ph.IM},
       adsurl = {https://ui.adsabs.harvard.edu/abs/2015A&C....10..121R}
}

@ARTICLE{hoekstra_17,
       author = {{Hoekstra}, Henk and {Viola}, Massimo and {Herbonnet}, Ricardo},
        title = "{A study of the sensitivity of shape measurements to the input parameters of weak-lensing image simulations}",
      journal = {\mnras},
         year = 2017,
        month = jul,
       volume = {468},
       number = {3},
        pages = {3295-3311},
          doi = {10.1093/mnras/stx724},
archivePrefix = {arXiv},
       eprint = {1609.03281},
 primaryClass = {astro-ph.CO},
       adsurl = {https://ui.adsabs.harvard.edu/abs/2017MNRAS.468.3295H}
}

@ARTICLE{sheldon_12,
       author = {{Sheldon}, Erin S. and {Cunha}, Carlos E. and {Mandelbaum}, Rachel and {Brinkmann}, J. and {Weaver}, Benjamin A.},
        title = "{Photometric Redshift Probability Distributions for Galaxies in the SDSS DR8}",
      journal = {\apjs},
         year = 2012,
        month = aug,
       volume = {201},
       number = {2},
          eid = {32},
        pages = {32},
          doi = {10.1088/0067-0049/201/2/32},
archivePrefix = {arXiv},
       eprint = {1109.5192},
 primaryClass = {astro-ph.CO},
       adsurl = {https://ui.adsabs.harvard.edu/abs/2012ApJS..201...32S}
}

@ARTICLE{deepdisc,
       author = {{Merz}, Grant and {Liu}, Yichen and {Burke}, Colin J. and {Aleo}, Patrick D. and {Liu}, Xin and {Carrasco Kind}, Matias and {Kindratenko}, Volodymyr and {Liu}, Yufeng},
        title = "{Detection, instance segmentation, and classification for astronomical surveys with deep learning (DEEPDISC): DETECTRON2 implementation and demonstration with Hyper Suprime-Cam data}",
      journal = {\mnras},
         year = 2023,
        month = nov,
       volume = {526},
       number = {1},
        pages = {1122-1137},
          doi = {10.1093/mnras/stad2785},
archivePrefix = {arXiv},
       eprint = {2307.05826},
 primaryClass = {astro-ph.IM},
       adsurl = {https://ui.adsabs.harvard.edu/abs/2023MNRAS.526.1122M}
}

@ARTICLE{cosmoDC2,
       author = {{Korytov}, Danila and {Hearin}, Andrew and {Kovacs}, Eve and {Larsen}, Patricia and {Rangel}, Esteban and {Hollowed}, Joseph and {Benson}, Andrew J. and {Heitmann}, Katrin and {Mao}, Yao-Yuan and {Bahmanyar}, Anita and {Chang}, Chihway and {Campbell}, Duncan and {DeRose}, Joseph and {Finkel}, Hal and {Frontiere}, Nicholas and {Gawiser}, Eric and {Habib}, Salman and {Joachimi}, Benjamin and {Lanusse}, Fran{\c{c}}ois and {Li}, Nan and {Mandelbaum}, Rachel and {Morrison}, Christopher and {Newman}, Jeffrey A. and {Pope}, Adrian and {Rykoff}, Eli and {Simet}, Melanie and {To}, Chun-Hao and {Vikraman}, Vinu and {Wechsler}, Risa H. and {White}, Martin and {(The LSST Dark Energy Science Collaboration}},
        title = "{CosmoDC2: A Synthetic Sky Catalog for Dark Energy Science with LSST}",
      journal = {\apjs},
         year = 2019,
        month = dec,
       volume = {245},
       number = {2},
          eid = {26},
        pages = {26},
          doi = {10.3847/1538-4365/ab510c},
archivePrefix = {arXiv},
       eprint = {1907.06530},
 primaryClass = {astro-ph.CO},
       adsurl = {https://ui.adsabs.harvard.edu/abs/2019ApJS..245...26K}
}

@ARTICLE{ssim,
  author={Zhou Wang and Bovik, A.C. and Sheikh, H.R. and Simoncelli, E.P.},
  journal={IEEE Transactions on Image Processing}, 
  title={Image quality assessment: from error visibility to structural similarity}, 
  year={2004},
  volume={13},
  number={4},
  pages={600-612},
  doi={10.1109/TIP.2003.819861}}

@ARTICLE{lsst_2019,
       author = {{Ivezi{\'c}}, {\v{Z}}eljko and {Kahn}, Steven M. and {Tyson}, J. Anthony and {Abel}, Bob and others},
        title = "{LSST: From Science Drivers to Reference Design and Anticipated Data Products}",
      journal = {\apj},
         year = 2019,
        month = mar,
       volume = {873},
       number = {2},
          eid = {111},
        pages = {111},
          doi = {10.3847/1538-4357/ab042c},
archivePrefix = {arXiv},
       eprint = {0805.2366},
 primaryClass = {astro-ph},
       adsurl = {https://ui.adsabs.harvard.edu/abs/2019ApJ...873..111I}
}

@INPROCEEDINGS{bliss,
       author = {{Hansen}, Derek L. and {Mendoza}, Ismael and {Liu}, Runjing and {Pang}, Ziteng and {Zhao}, Zhe and {Avestruz}, Camille and {Regier}, Jeffrey},
        title = "{Scalable Bayesian Inference for Detection and Deblending in Astronomical Images}",
    booktitle = {Machine Learning for Astrophysics},
         year = 2022,
        month = jul,
          eid = {27},
        pages = {27},
          doi = {10.48550/arXiv.2207.05642},
archivePrefix = {arXiv},
       eprint = {2207.05642},
 primaryClass = {astro-ph.IM},
       adsurl = {https://ui.adsabs.harvard.edu/abs/2022mla..confE..27H}
}

@Article{madness,
  author        = {{Biswas}, B. and {Aubourg}, E. and {Boucaud}, A. and {Guinot}, A. and {Lao}, J. and {Roucelle}, C. and {LSST Dark Energy Science Collaboration}},
  journal       = {\aap},
  title         = {{MADNESS deblender: Maximum A posteriori with Deep NEural networks for Source Separation}},
  year          = {2025},
  month         = aug,
  pages         = {A129},
  volume        = {700},
  adsurl        = {https://ui.adsabs.harvard.edu/abs/2025A&A...700A.129B},
  archiveprefix = {arXiv},
  doi           = {10.1051/0004-6361/202451887},
  eid           = {A129},
  eprint        = {2408.15236},
  primaryclass  = {astro-ph.IM},
}

@ARTICLE{3x2pt_desi,
       author = {{Blake}, Chris and {Garcia-Quintero}, C. and {Ahlen}, S. and {Bianchi}, D. and {Brooks}, D. and {Claybaugh}, T. and {de la Macorra}, A. and {DeRose}, J. and {Dey}, A. and {Doel}, P. and {Emas}, N. and {Ferraro}, S. and {Forero-Romero}, J.~E. and {Gutierrez}, G. and {Heydenreich}, S. and {Honscheid}, K. and {Howlett}, C. and {Ishak}, M. and {Joudaki}, S. and {Jullo}, E. and {Kehoe}, R. and {Kirkby}, D. and {Kremin}, A. and {Krolewski}, A. and {Landriau}, M. and {Lange}, J.~U. and {Leauthaud}, A. and {Levi}, M.~E. and {Manera}, M. and {Miquel}, R. and {Moustakas}, J. and {Niz}, G. and {Percival}, W.~J. and {P{\'e}rez-R{\`a}fols}, I. and {Porredon}, A. and {Rossi}, G. and {Ruggeri}, R. and {Sanchez}, E. and {Saulder}, C. and {Schlegel}, D. and {Sprayberry}, D. and {Sun}, Z. and {Tarl{\'e}}, G. and {Weaver}, B.~A.},
        title = "{The DESI-Lensing Mock Challenge: large-scale cosmological analysis of 3x2-pt statistics}",
      journal = {The Open Journal of Astrophysics},
         year = 2025,
        month = mar,
       volume = {8},
          eid = {24},
        pages = {24},
          doi = {10.33232/001c.131903},
archivePrefix = {arXiv},
       eprint = {2412.12548},
 primaryClass = {astro-ph.CO},
       adsurl = {https://ui.adsabs.harvard.edu/abs/2025OJAp....8E..24B}
}

@ARTICLE{Nourbakhsh_22,
       author = {{Nourbakhsh}, Erfan and {Tyson}, J. Anthony and {Schmidt}, Samuel J. and {LSST Dark Energy Science Collaboration}},
        title = "{Galaxy blending effects in deep imaging cosmic shear probes of cosmology}",
      journal = {\mnras},
         year = 2022,
        month = aug,
       volume = {514},
       number = {4},
        pages = {5905-5926},
          doi = {10.1093/mnras/stac1303},
archivePrefix = {arXiv},
       eprint = {2112.07659},
 primaryClass = {astro-ph.CO},
       adsurl = {https://ui.adsabs.harvard.edu/abs/2022MNRAS.514.5905N}
}

@ARTICLE{desc_ai_2026_WHITE_PAPER,
       author = {{LSST Dark Energy Science Collaboration} and {Aubourg}, Eric and {Avestruz}, Camille and {Becker}, Matthew R. and {Biswas}, Biswajit and {Biswas}, Rahul and {Bolliet}, Boris and {Bolton}, Adam S. and {Bom}, Clecio R. and {Bonnet-Guerrini}, Rapha{\"e}l and {Boucaud}, Alexandre and {Campagne}, Jean-Eric and {Chang}, Chihway and {{\'C}iprijanovi{\'c}}, Aleksandra and {Cohen-Tanugi}, Johann and {Coughlin}, Michael W. and {Crenshaw}, John Franklin and {Cuevas-Tello}, Juan C. and {de Vicente}, Juan and {Digel}, Seth W. and {Dillmann}, Steven and {de Le{\'o}n Dominguez Romero}, Mariano Javier and {Drlica-Wagner}, Alex and {Erickson}, Sydney and {Gagliano}, Alexander T. and {Georgiou}, Christos and {Ghosh}, Aritra and {Grayling}, Matthew and {Grishin}, Kirill A. and {Heavens}, Alan and {House}, Lindsay R. and {Ishak}, Mustapha and {Kabalan}, Wassim and {Kannawadi}, Arun and {Lanusse}, Fran{\c{c}}ois and {Leonard}, C. Danielle and {L{\'e}get}, Pierre-Fran{\c{c}}ois and {Lochner}, Michelle and {Mao}, Yao-Yuan and {Melchior}, Peter and {Merz}, Grant and {Millon}, Martin and {M{\"o}ller}, Anais and {Narayan}, Gautham and {Omori}, Yuuki and {Peiris}, Hiranya and {Perreault-Levasseur}, Laurence and {Plazas Malag{\'o}n}, Andr{\'e}s A. and {Ramachandra}, Nesar and {Remy}, Benjamin and {Roucelle}, C{\'e}cile and {Ruiz-Zapatero}, Jaime and {Schuldt}, Stefan and {Sevilla-Noarbe}, Ignacio and {Shah}, Ved G. and {Starkenburg}, Tjitske and {Thorp}, Stephen and {Toribio San Cipriano}, Laura and {Tr{\"o}ster}, Tilman and {Trotta}, Roberto and {Venkatraman}, Padma and {Wasserman}, Amanda and {White}, Tim and {Zeghal}, Justine and {Zhang}, Tianqing and {Zhang}, Yuanyuan},
        title = "{Opportunities in AI/ML for the Rubin LSST Dark Energy Science Collaboration}",
      journal = {arXiv e-prints},
         year = 2026,
        month = jan,
          eid = {arXiv:2601.14235},
        pages = {arXiv:2601.14235},
          doi = {10.48550/arXiv.2601.14235},
archivePrefix = {arXiv},
       eprint = {2601.14235},
 primaryClass = {astro-ph.IM},
       adsurl = {https://ui.adsabs.harvard.edu/abs/2026arXiv260114235L}
}

@ARTICLE{deepdisc_photoz,
       author = {{Merz}, Grant and {Liu}, Xin and {Schmidt}, Samuel and {Malz}, Alex I. and {Zhang}, Tianqing and {Branton}, Doug and {Burke}, Colin J. and {Delucchi}, Melissa and {Ejjagiri}, Yaswant Sai and {Kubica}, Jeremy and {Liu}, Yichen and {Lynn}, Olivia and {Oldag}, Drew and {LSST Dark Energy Science Collaboration}},
        title = "{DeepDISC-photoz: Deep Learning-Based Photometric Redshift Estimation for Rubin LSST}",
      journal = {The Open Journal of Astrophysics},
         year = 2025,
        month = apr,
       volume = {8},
          eid = {40},
        pages = {40},
          doi = {10.33232/001c.136809},
archivePrefix = {arXiv},
       eprint = {2411.18769},
 primaryClass = {astro-ph.IM},
       adsurl = {https://ui.adsabs.harvard.edu/abs/2025OJAp....8E..40M}
}

@TechReport{lsst_science_pipelines,
   author = "{Vera C. Rubin Observatory Science Pipelines Developers}",
   title = "{The LSST Science Pipelines Software: Optical Survey Pipeline Reduction and Analysis Environment}",
   institution = "{Vera C. Rubin Observatory}",
   year = "2025",
   month = "June",
   handle = "PSTN-019",
   type = "{Project Science Technical Note}",
   number = "PSTN-019",
   doi = "10.71929/rubin/2570545",
   url = "https://pstn-019.lsst.io/"
}

@ARTICLE{molino_2017,
       author = {{Molino}, A. and {Ben{\'\i}tez}, N. and {Ascaso}, B. and {Coe}, D. and {Postman}, M. and {Jouvel}, S. and {Host}, O. and {Lahav}, O. and {Seitz}, S. and {Medezinski}, E. and {Rosati}, P. and {Schoenell}, W. and {Koekemoer}, A. and {Jimenez-Teja}, Y. and {Broadhurst}, T. and {Melchior}, P. and {Balestra}, I. and {Bartelmann}, M. and {Bouwens}, R. and {Bradley}, L. and {Czakon}, N. and {Donahue}, M. and {Ford}, H. and {Graur}, O. and {Graves}, G. and {Grillo}, C. and {Infante}, L. and {Jha}, S.~W. and {Kelson}, D. and {Lazkoz}, R. and {Lemze}, D. and {Maoz}, D. and {Mercurio}, A. and {Meneghetti}, M. and {Merten}, J. and {Moustakas}, L. and {Nonino}, M. and {Orgaz}, S. and {Riess}, A. and {Rodney}, S. and {Sayers}, J. and {Umetsu}, K. and {Zheng}, W. and {Zitrin}, A.},
        title = "{CLASH: accurate photometric redshifts with 14 HST bands in massive galaxy cluster cores}",
      journal = {\mnras},
         year = 2017,
        month = sep,
       volume = {470},
       number = {1},
        pages = {95-113},
          doi = {10.1093/mnras/stx1243},
archivePrefix = {arXiv},
       eprint = {1705.02265},
 primaryClass = {astro-ph.GA},
       adsurl = {https://ui.adsabs.harvard.edu/abs/2017MNRAS.470...95M}
}

@ARTICLE{melchior_2021,
       author = {{Melchior}, Peter and {Joseph}, R{\'e}my and {Sanchez}, Javier and {MacCrann}, Niall and {Gruen}, Daniel},
        title = "{The challenge of blending in large sky surveys}",
      journal = {Nature Reviews Physics},
         year = 2021,
        month = aug,
       volume = {3},
       number = {10},
        pages = {712-718},
          doi = {10.1038/s42254-021-00353-y},
       adsurl = {https://ui.adsabs.harvard.edu/abs/2021NatRP...3..712M}
}

@ARTICLE{sanchez_overlap_cosmic_shear,
       author = {{Sanchez}, Javier and {Mendoza}, Ismael and {Kirkby}, David P. and {Burchat}, Patricia R. and {LSST Dark Energy Science Collaboration}},
        title = "{Effects of overlapping sources on cosmic shear estimation: Statistical sensitivity and pixel-noise bias}",
      journal = {\jcap},
         year = 2021,
        month = jul,
       volume = {2021},
       number = {7},
          eid = {043},
        pages = {043},
          doi = {10.1088/1475-7516/2021/07/043},
archivePrefix = {arXiv},
       eprint = {2103.02078},
 primaryClass = {astro-ph.CO},
       adsurl = {https://ui.adsabs.harvard.edu/abs/2021JCAP...07..043S}
}

@ARTICLE{des_yr1_photoz,
       author = {{Gruen}, D. and {Zhang}, Y. and {Palmese}, A. and {Yanny}, B. and {Busti}, V. and {Hoyle}, B. and {Melchior}, P. and {Miller}, C.~J. and {Rozo}, E. and {Rykoff}, E.~S. and {Varga}, T.~N. and {Abdalla}, F.~B. and {Allam}, S. and {Annis}, J. and {Avila}, S. and {Brooks}, D. and {Burke}, D.~L. and {Carnero Rosell}, A. and {Carrasco Kind}, M. and {Carretero}, J. and {Cawthon}, R. and {Crocce}, M. and {Cunha}, C.~E. and {da Costa}, L.~N. and {Davis}, C. and {De Vicente}, J. and {Desai}, S. and {Diehl}, H.~T. and {Dietrich}, J.~P. and {Drlica-Wagner}, A. and {Flaugher}, B. and {Fosalba}, P. and {Frieman}, J. and {Garc{\'\i}a-Bellido}, J. and {Gaztanaga}, E. and {Gerdes}, D.~W. and {Gruendl}, R.~A. and {Gschwend}, J. and {Hollowood}, D.~L. and {Honscheid}, K. and {James}, D.~J. and {Jeltema}, T. and {Krause}, E. and {Kron}, R. and {Kuehn}, K. and {Kuropatkin}, N. and {Lahav}, O. and {Lima}, M. and {Lin}, H. and {Maia}, M.~A.~G. and {Marshall}, J.~L. and {Menanteau}, F. and {Miquel}, R. and {Ogando}, R.~L.~C. and {Plazas}, A.~A. and {Romer}, A.~K. and {Scarpine}, V. and {Sevilla-Noarbe}, I. and {Smith}, M. and {Soares-Santos}, M. and {Sobreira}, F. and {Suchyta}, E. and {Swanson}, M.~E.~C. and {Tarle}, G. and {Thomas}, D. and {Vikram}, V. and {Walker}, A.~R. and {DES Collaboration}},
        title = "{Dark Energy Survey Year 1 results: the effect of intracluster light on photometric redshifts for weak gravitational lensing}",
      journal = {\mnras},
         year = 2019,
        month = sep,
       volume = {488},
       number = {3},
        pages = {4389-4399},
          doi = {10.1093/mnras/stz2036},
archivePrefix = {arXiv},
       eprint = {1809.04599},
 primaryClass = {astro-ph.CO},
       adsurl = {https://ui.adsabs.harvard.edu/abs/2019MNRAS.488.4389G}
}

@ARTICLE{samuroff_2018,
       author = {{Samuroff}, S. and {Bridle}, S.~L. and {Zuntz}, J. and {Troxel}, M.~A. and {Gruen}, D. and {Rollins}, R.~P. and {Bernstein}, G.~M. and {Eifler}, T.~F. and {Huff}, E.~M. and {Kacprzak}, T. and {Krause}, E. and {MacCrann}, N. and {Abdalla}, F.~B. and {Allam}, S. and {Annis}, J. and {Bechtol}, K. and {Benoit-L{\'e}vy}, A. and {Bertin}, E. and {Brooks}, D. and {Buckley-Geer}, E. and {Carnero Rosell}, A. and {Carrasco Kind}, M. and {Carretero}, J. and {Crocce}, M. and {D'Andrea}, C.~B. and {da Costa}, L.~N. and {Davis}, C. and {Desai}, S. and {Doel}, P. and {Fausti Neto}, A. and {Flaugher}, B. and {Fosalba}, P. and {Frieman}, J. and {Garc{\'\i}a-Bellido}, J. and {Gerdes}, D.~W. and {Gruendl}, R.~A. and {Gschwend}, J. and {Gutierrez}, G. and {Honscheid}, K. and {James}, D.~J. and {Jarvis}, M. and {Jeltema}, T. and {Kirk}, D. and {Kuehn}, K. and {Kuhlmann}, S. and {Li}, T.~S. and {Lima}, M. and {Maia}, M.~A.~G. and {March}, M. and {Marshall}, J.~L. and {Martini}, P. and {Melchior}, P. and {Menanteau}, F. and {Miquel}, R. and {Nord}, B. and {Ogando}, R.~L.~C. and {Plazas}, A.~A. and {Roodman}, A. and {Sanchez}, E. and {Scarpine}, V. and {Schindler}, R. and {Schubnell}, M. and {Sevilla-Noarbe}, I. and {Sheldon}, E. and {Smith}, M. and {Soares-Santos}, M. and {Sobreira}, F. and {Suchyta}, E. and {Tarle}, G. and {Thomas}, D. and {Tucker}, D.~L. and {DES Collaboration}},
        title = "{Dark Energy Survey Year 1 results: the impact of galaxy neighbours on weak lensing cosmology with IM3SHAPE}",
      journal = {\mnras},
         year = 2018,
        month = apr,
       volume = {475},
       number = {4},
        pages = {4524-4543},
          doi = {10.1093/mnras/stx3282},
archivePrefix = {arXiv},
       eprint = {1708.01534},
 primaryClass = {astro-ph.CO},
       adsurl = {https://ui.adsabs.harvard.edu/abs/2018MNRAS.475.4524S}
}

@ARTICLE{HSC,
       author = {{Aihara}, Hiroaki and {Arimoto}, Nobuo and {Armstrong}, Robert and {Arnouts}, St{\'e}phane and {Bahcall}, Neta A. and {Bickerton}, Steven and {Bosch}, James and {Bundy}, Kevin and {Capak}, Peter L. and {Chan}, James H.~H. and {Chiba}, Masashi and {Coupon}, Jean and {Egami}, Eiichi and {Enoki}, Motohiro and {Finet}, Francois and {Fujimori}, Hiroki and {Fujimoto}, Seiji and {Furusawa}, Hisanori and {Furusawa}, Junko and {Goto}, Tomotsugu and {Goulding}, Andy and {Greco}, Johnny P. and {Greene}, Jenny E. and {Gunn}, James E. and {Hamana}, Takashi and {Harikane}, Yuichi and {Hashimoto}, Yasuhiro and {Hattori}, Takashi and {Hayashi}, Masao and {Hayashi}, Yusuke and {He{\l}miniak}, Krzysztof G. and {Higuchi}, Ryo and {Hikage}, Chiaki and {Ho}, Paul T.~P. and {Hsieh}, Bau-Ching and {Huang}, Kuiyun and {Huang}, Song and {Ikeda}, Hiroyuki and {Imanishi}, Masatoshi and {Inoue}, Akio K. and {Iwasawa}, Kazushi and {Iwata}, Ikuru and {Jaelani}, Anton T. and {Jian}, Hung-Yu and {Kamata}, Yukiko and {Karoji}, Hiroshi and {Kashikawa}, Nobunari and {Katayama}, Nobuhiko and {Kawanomoto}, Satoshi and {Kayo}, Issha and {Koda}, Jin and {Koike}, Michitaro and {Kojima}, Takashi and {Komiyama}, Yutaka and {Konno}, Akira and {Koshida}, Shintaro and {Koyama}, Yusei and {Kusakabe}, Haruka and {Leauthaud}, Alexie and {Lee}, Chien-Hsiu and {Lin}, Lihwai and {Lin}, Yen-Ting and {Lupton}, Robert H. and {Mandelbaum}, Rachel and {Matsuoka}, Yoshiki and {Medezinski}, Elinor and {Mineo}, Sogo and {Miyama}, Shoken and {Miyatake}, Hironao and {Miyazaki}, Satoshi and {Momose}, Rieko and {More}, Anupreeta and {More}, Surhud and {Moritani}, Yuki and {Moriya}, Takashi J. and {Morokuma}, Tomoki and {Mukae}, Shiro and {Murata}, Ryoma and {Murayama}, Hitoshi and {Nagao}, Tohru and {Nakata}, Fumiaki and {Niida}, Mana and {Niikura}, Hiroko and {Nishizawa}, Atsushi J. and {Obuchi}, Yoshiyuki and {Oguri}, Masamune and {Oishi}, Yukie and {Okabe}, Nobuhiro and {Okamoto}, Sakurako and {Okura}, Yuki and {Ono}, Yoshiaki and {Onodera}, Masato and {Onoue}, Masafusa and {Osato}, Ken and {Ouchi}, Masami and {Price}, Paul A. and {Pyo}, Tae-Soo and {Sako}, Masao and {Sawicki}, Marcin and {Shibuya}, Takatoshi and {Shimasaku}, Kazuhiro and {Shimono}, Atsushi and {Shirasaki}, Masato and {Silverman}, John D. and {Simet}, Melanie and {Speagle}, Joshua and {Spergel}, David N. and {Strauss}, Michael A. and {Sugahara}, Yuma and {Sugiyama}, Naoshi and {Suto}, Yasushi and {Suyu}, Sherry H. and {Suzuki}, Nao and {Tait}, Philip J. and {Takada}, Masahiro and {Takata}, Tadafumi and {Tamura}, Naoyuki and {Tanaka}, Manobu M. and {Tanaka}, Masaomi and {Tanaka}, Masayuki and {Tanaka}, Yoko and {Terai}, Tsuyoshi and {Terashima}, Yuichi and {Toba}, Yoshiki and {Tominaga}, Nozomu and {Toshikawa}, Jun and {Turner}, Edwin L. and {Uchida}, Tomohisa and {Uchiyama}, Hisakazu and {Umetsu}, Keiichi and {Uraguchi}, Fumihiro and {Urata}, Yuji and {Usuda}, Tomonori and {Utsumi}, Yousuke and {Wang}, Shiang-Yu and {Wang}, Wei-Hao and {Wong}, Kenneth C. and {Yabe}, Kiyoto and {Yamada}, Yoshihiko and {Yamanoi}, Hitomi and {Yasuda}, Naoki and {Yeh}, Sherry and {Yonehara}, Atsunori and {Yuma}, Suraphong},
        title = "{The Hyper Suprime-Cam SSP Survey: Overview and survey design}",
      journal = {\pasj},
         year = 2018,
        month = jan,
       volume = {70},
          eid = {S4},
        pages = {S4},
          doi = {10.1093/pasj/psx066},
archivePrefix = {arXiv},
       eprint = {1704.05858},
 primaryClass = {astro-ph.IM},
       adsurl = {https://ui.adsabs.harvard.edu/abs/2018PASJ...70S...4A}
}

@ARTICLE{hsc_pipeline,
       author = {{Bosch}, James and {Armstrong}, Robert and {Bickerton}, Steven and {Furusawa}, Hisanori and {Ikeda}, Hiroyuki and {Koike}, Michitaro and {Lupton}, Robert and {Mineo}, Sogo and {Price}, Paul and {Takata}, Tadafumi and {Tanaka}, Masayuki and {Yasuda}, Naoki and {AlSayyad}, Yusra and {Becker}, Andrew C. and {Coulton}, William and {Coupon}, Jean and {Garmilla}, Jose and {Huang}, Song and {Krughoff}, K. Simon and {Lang}, Dustin and {Leauthaud}, Alexie and {Lim}, Kian-Tat and {Lust}, Nate B. and {MacArthur}, Lauren A. and {Mandelbaum}, Rachel and {Miyatake}, Hironao and {Miyazaki}, Satoshi and {Murata}, Ryoma and {More}, Surhud and {Okura}, Yuki and {Owen}, Russell and {Swinbank}, John D. and {Strauss}, Michael A. and {Yamada}, Yoshihiko and {Yamanoi}, Hitomi},
        title = "{The Hyper Suprime-Cam software pipeline}",
      journal = {\pasj},
         year = 2018,
        month = jan,
       volume = {70},
          eid = {S5},
        pages = {S5},
          doi = {10.1093/pasj/psx080},
archivePrefix = {arXiv},
       eprint = {1705.06766},
 primaryClass = {astro-ph.IM},
       adsurl = {https://ui.adsabs.harvard.edu/abs/2018PASJ...70S...5B}
}

@ARTICLE{DES,
       author = {{Dark Energy Survey Collaboration} and {Abbott}, T. and {Abdalla}, F.~B. and {Aleksi{\'c}}, J. and {Allam}, S. and {Amara}, A. and {Bacon}, D. and {Balbinot}, E. and {Banerji}, M. and {Bechtol}, K. and {Benoit-L{\'e}vy}, A. and {Bernstein}, G.~M. and {Bertin}, E. and {Blazek}, J. and {Bonnett}, C. and {Bridle}, S. and {Brooks}, D. and {Brunner}, R.~J. and {Buckley-Geer}, E. and {Burke}, D.~L. and {Caminha}, G.~B. and {Capozzi}, D. and {Carlsen}, J. and {Carnero-Rosell}, A. and {Carollo}, M. and {Carrasco-Kind}, M. and {Carretero}, J. and {Castander}, F.~J. and {Clerkin}, L. and {Collett}, T. and {Conselice}, C. and {Crocce}, M. and {Cunha}, C.~E. and {D'Andrea}, C.~B. and {da Costa}, L.~N. and {Davis}, T.~M. and {Desai}, S. and {Diehl}, H.~T. and {Dietrich}, J.~P. and {Dodelson}, S. and {Doel}, P. and {Drlica-Wagner}, A. and {Estrada}, J. and {Etherington}, J. and {Evrard}, A.~E. and {Fabbri}, J. and {Finley}, D.~A. and {Flaugher}, B. and {Foley}, R.~J. and {Fosalba}, P. and {Frieman}, J. and {Garc{\'\i}a-Bellido}, J. and {Gaztanaga}, E. and {Gerdes}, D.~W. and {Giannantonio}, T. and {Goldstein}, D.~A. and {Gruen}, D. and {Gruendl}, R.~A. and {Guarnieri}, P. and {Gutierrez}, G. and {Hartley}, W. and {Honscheid}, K. and {Jain}, B. and {James}, D.~J. and {Jeltema}, T. and {Jouvel}, S. and {Kessler}, R. and {King}, A. and {Kirk}, D. and {Kron}, R. and {Kuehn}, K. and {Kuropatkin}, N. and {Lahav}, O. and {Li}, T.~S. and {Lima}, M. and {Lin}, H. and {Maia}, M.~A.~G. and {Makler}, M. and {Manera}, M. and {Maraston}, C. and {Marshall}, J.~L. and {Martini}, P. and {McMahon}, R.~G. and {Melchior}, P. and {Merson}, A. and {Miller}, C.~J. and {Miquel}, R. and {Mohr}, J.~J. and {Morice-Atkinson}, X. and {Naidoo}, K. and {Neilsen}, E. and {Nichol}, R.~C. and {Nord}, B. and {Ogando}, R. and {Ostrovski}, F. and {Palmese}, A. and {Papadopoulos}, A. and {Peiris}, H.~V. and {Peoples}, J. and {Percival}, W.~J. and {Plazas}, A.~A. and {Reed}, S.~L. and {Refregier}, A. and {Romer}, A.~K. and {Roodman}, A. and {Ross}, A. and {Rozo}, E. and {Rykoff}, E.~S. and {Sadeh}, I. and {Sako}, M. and {S{\'a}nchez}, C. and {Sanchez}, E. and {Santiago}, B. and {Scarpine}, V. and {Schubnell}, M. and {Sevilla-Noarbe}, I. and {Sheldon}, E. and {Smith}, M. and {Smith}, R.~C. and {Soares-Santos}, M. and {Sobreira}, F. and {Soumagnac}, M. and {Suchyta}, E. and {Sullivan}, M. and {Swanson}, M. and {Tarle}, G. and {Thaler}, J. and {Thomas}, D. and {Thomas}, R.~C. and {Tucker}, D. and {Vieira}, J.~D. and {Vikram}, V. and {Walker}, A.~R. and {Wechsler}, R.~H. and {Weller}, J. and {Wester}, W. and {Whiteway}, L. and {Wilcox}, H. and {Yanny}, B. and {Zhang}, Y. and {Zuntz}, J.},
        title = "{The Dark Energy Survey: more than dark energy - an overview}",
      journal = {\mnras},
         year = 2016,
        month = aug,
       volume = {460},
       number = {2},
        pages = {1270-1299},
          doi = {10.1093/mnras/stw641},
archivePrefix = {arXiv},
       eprint = {1601.00329},
 primaryClass = {astro-ph.CO},
       adsurl = {https://ui.adsabs.harvard.edu/abs/2016MNRAS.460.1270D}
}

@ARTICLE{vae_reconstructions,
       author = {{Arcelin}, Bastien and {Doux}, Cyrille and {Aubourg}, Eric and {Roucelle}, C{\'e}cile and {LSST Dark Energy Science Collaboration}},
        title = "{Deblending galaxies with variational autoencoders: A joint multiband, multi-instrument approach}",
      journal = {\mnras},
         year = 2021,
        month = jan,
       volume = {500},
       number = {1},
        pages = {531-547},
          doi = {10.1093/mnras/staa3062},
archivePrefix = {arXiv},
       eprint = {2005.12039},
 primaryClass = {astro-ph.IM},
       adsurl = {https://ui.adsabs.harvard.edu/abs/2021MNRAS.500..531A}
}

@ARTICLE{dc2_creation,
       author = {{LSST Dark Energy Science Collaboration (LSST DESC)} and {Abolfathi}, Bela and {Alonso}, David and {Armstrong}, Robert and {Aubourg}, {\'E}ric and {Awan}, Humna and {Babuji}, Yadu N. and {Bauer}, Franz Erik and {Bean}, Rachel and {Beckett}, George and {Biswas}, Rahul and {Bogart}, Joanne R. and {Boutigny}, Dominique and {Chard}, Kyle and {Chiang}, James and {Claver}, Chuck F. and {Cohen-Tanugi}, Johann and {Combet}, C{\'e}line and {Connolly}, Andrew J. and {Daniel}, Scott F. and {Digel}, Seth W. and {Drlica-Wagner}, Alex and {Dubois}, Richard and {Gangler}, Emmanuel and {Gawiser}, Eric and {Glanzman}, Thomas and {Gris}, Phillipe and {Habib}, Salman and {Hearin}, Andrew P. and {Heitmann}, Katrin and {Hernandez}, Fabio and {Hlo{\v{z}}ek}, Ren{\'e}e and {Hollowed}, Joseph and {Ishak}, Mustapha and {Ivezi{\'c}}, {\v{Z}}eljko and {Jarvis}, Mike and {Jha}, Saurabh W. and {Kahn}, Steven M. and {Kalmbach}, J. Bryce and {Kelly}, Heather M. and {Kovacs}, Eve and {Korytov}, Danila and {Krughoff}, K. Simon and {Lage}, Craig S. and {Lanusse}, Fran{\c{c}}ois and {Larsen}, Patricia and {Le Guillou}, Laurent and {Li}, Nan and {Longley}, Emily Phillips and {Lupton}, Robert H. and {Mandelbaum}, Rachel and {Mao}, Yao-Yuan and {Marshall}, Phil and {Meyers}, Joshua E. and {Moniez}, Marc and {Morrison}, Christopher B. and {Nomerotski}, Andrei and {O'Connor}, Paul and {Park}, HyeYun and {Park}, Ji Won and {Peloton}, Julien and {Perrefort}, Daniel and {Perry}, James and {Plaszczynski}, St{\'e}phane and {Pope}, Adrian and {Rasmussen}, Andrew and {Reil}, Kevin and {Roodman}, Aaron J. and {Rykoff}, Eli S. and {S{\'a}nchez}, F. Javier and {Schmidt}, Samuel J. and {Scolnic}, Daniel and {Stubbs}, Christopher W. and {Tyson}, J. Anthony and {Uram}, Thomas D. and {Villarreal}, Antonia Sierra and {Walter}, Christopher W. and {Wiesner}, Matthew P. and {Wood-Vasey}, W. Michael and {Zuntz}, Joe},
        title = "{The LSST DESC DC2 Simulated Sky Survey}",
      journal = {\apjs},
         year = 2021,
        month = mar,
       volume = {253},
       number = {1},
          eid = {31},
        pages = {31},
          doi = {10.3847/1538-4365/abd62c},
archivePrefix = {arXiv},
       eprint = {2010.05926},
 primaryClass = {astro-ph.IM},
       adsurl = {https://ui.adsabs.harvard.edu/abs/2021ApJS..253...31L}
}

@ARTICLE{universemachine,
       author = {{Behroozi}, Peter and {Wechsler}, Risa H. and {Hearin}, Andrew P. and {Conroy}, Charlie},
        title = "{UNIVERSEMACHINE: The correlation between galaxy growth and dark matter halo assembly from z = 0-10}",
      journal = {\mnras},
         year = 2019,
        month = sep,
       volume = {488},
       number = {3},
        pages = {3143-3194},
          doi = {10.1093/mnras/stz1182},
archivePrefix = {arXiv},
       eprint = {1806.07893},
 primaryClass = {astro-ph.GA},
       adsurl = {https://ui.adsabs.harvard.edu/abs/2019MNRAS.488.3143B}
}

@misc{wu2019detectron2,
  author =       {Yuxin Wu and Alexander Kirillov and Francisco Massa and
                  Wan-Yen Lo and Ross Girshick},
  title =        {Detectron2},
  howpublished = {\url{https://github.com/facebookresearch/detectron2}},
  year =         {2019}
}

@Article{LSSTDP1,
  author        = {{Vera C Rubin Observatory Team} and Cuellar, Tatiana Acero and Acosta, Emily and Adair, Christina L and Adari, Prakruth and McCarthy, Jennifer K Adelman and Alexov, Anastasia and Allbery, Russ and Allsman, Robyn and AlSayyad, Yusra and Amado, Jhonatan and Amouroux, Nathan and Antilogus, Pierre and Alcayaga, Alexis Aracena and Rojas, Gonzalo Aravena and Cortes, Claudio H Araya and Aubourg, Eric and Axelrod, Tim S and Banovetz, John and Barria, Carlos and Bauer, Amanda E and Bauman, Brian J and Bechtol, Ellen and Bechtol, Keith and Becker, Andrew C and Becker, Valerie R and Beckett, Mark G and Bellm, Eric C and Bernardinelli, Pedro H and Bianco, Federica Bettina and Blum, Robert D and Bogart, Joanne and Bolton, Adam and Booth, Michael T and Bosch, James F and Boucaud, Alexandre and Boutigny, Dominique and Bovill, Robert A and Bradshaw, Andrew and Bregeon, Johan and Brescia, Massimo and Brondel, Brian J and Broughton, Alexander and Budlong, Audrey and Buffat, Dimitri and Canestrari, Rodolfo and Caplar, Neven and Carlin, Jeffrey L and Ceballo, Ross and Chandler, Colin Orion and Chang, Chihway and Emerson, Glenaver Charles and Chiang, Hsin Fang and Chiang, James and Choi, Yumi and Christensen, Eric J and Claver, Charles F and Clements, Andy W and Cockrum, Joseph J and Tanugi, Johann Cohen and Colleoni, Franco and Combet, Celine and Connolly, Andrew J and Cordova, Julio Eduardo Constanzo and Contreras, Hans E and Crenshaw, John Franklin and Campagne, Sylvie Dagoret and Daniel, Scott F and Daruich, Felipe and Daubard, Guillaume and Daues, Greg and Dennihy, Erik and Deppe, Stephanie J H and Digel, Seth W and Doherty, Peter E and Doux, Cyrille and Wagner, Alex Drlica and Felsmann, Gregory P Dubois and Economou, Frossie and Eiger, Orion and Eisert, Lukas and Eisner, Alan M and Englert, Anthony and Erb, Baden and Fabrega, Juan A and Fagrelius, Parker and Fanning, Kevin and Neto, Angelo Fausti and Ferguson, Peter S and Ferte, Agnes and Findeisen, Krzysztof and Levine, Merlin Fisher and Alvarez, Gloria Fonseca and Foss, Michael D and Fouchez, Dominique and Fuchs, Dan C and Fu, Shenming and Gangler, Emmanuel and Gaponenko, Igor and Garcia, Julen and Gates, John H and Gill, Ranpal K and Giro, Enrico and Glanzman, Thomas and Godoy, Robinson and Goodenow, Iain and Gorsuch, Miranda R and Gower, Michelle and Graham, Melissa L and Granvik, Mikael and Greenstreet, Sarah and Guan, Wen and Guillemin, Thibault and Guy, Leanne P and Hascall, Diane and Hascall, Patrick A and Heinze, Aren Nathaniel and Hernandez, Fabio and Herner, Kenneth and Herrold, Ardis and Higgs, Clare R and Hoblitt, Joshua and Howard, Erin Leigh and Hyun, Minhee and Ibsen, Amanda and Ingraham, Patrick and Irving, David H and Ivezic, Zeljko and Jacoby, Suzanne H and Jannuzi, Buell T and Jarugula, Sreevani and Jee, M James and Jenness, Tim and Jennings, Toby C and Jeremie, Andrea and Jernigan, Garrett and Mejias, David Jimenez and Johnson, Anthony S and Jones, R Lynne and Jones, Roger William Lewis and Gilles, Claire Juramy and Juric, Mario and Kahn, Steven M and Kalmbach, J Bryce and Kang, Yijung and Kannawadi, Arun and Kantor, Jeffrey P and Karavakis, Edward and Kelkar, Kshitija and Kelvin, Lee S and Kleinman, Scot J and Kotov, Ivan V and Kovacs, Gabor and Kowalik, Mikolaj and Krabbendam, Victor L and Krughoff, K Simon and Kubanek, Petr and Kurlander, Jacob A and Kusulja, Mile and Lage, Craig S and Lago, Paulo J A and Laliotis, Katherine and Lange, Travis and Laporte, Didier and Lau, Ryan M and Lazarte, Juan Carlos and Boulch, Quentin Le and Leget, Pierre Francois and Guillou, Laurent Le and Levine, Benjamin and Liang, Ming and Liang, Shuang and Lim, Kian Tat and von der Linden, Anja and Lin, Huan and Lopez, Margaux and Toro, Juan J Lopez and Love, Peter and Lupton, Robert H and Lust, Nate B and MacArthur, Lauren A and MacBride, Sean Patrick and Madejski, Greg M and Mainetti, Gabriele and Margheim, Steven J and Markiewicz, Thomas W and Marshall, Phil and Marshall, Stuart and Maulen, Guido and Mau, Sidney and May, Morgan and McCormick, Jeremy and McKay, David and McKercher, Robert and Homar, Guillem Megias and Meisner, Aaron M and Menanteau, Felipe and Mentzer, Heather R and Metzger, Kristen and Meyers, Joshua E and Miller, Michelle and Mills, David J and Moeyens, Joachim and Moniez, Marc and Moolekamp, Fred E and Marin, C A L Morales and Mueller, Fritz and Mullaney, James R and Arancibia, Freddy Munoz and Napier, Kate and Neal, Homer and Neilsen, Eric H and Neveu, Jeremy and Noble, Timothy and Nourbakhsh, Erfan and Olsen, Knut and OMullane, William and Onoprienko, Dmitry and Oriunno, Marco and Osier, Shawn and Owen, Russell E and Pai, Aashay and Parejko, John K and Park, Hye Yun and Parsons, James B and Patterson, Maria T and Pavlovic, Marina S and Ramirez, Karla Pena and Peterson, John R and Pietrowicz, Stephen R and Malagon, Andres A Plazas and Polen, Rebekah and Pollek, Hannah Mary Margaret and Price, Paul A and Quint, Bruno C and Marin, Jose Miguel Quintero and Rabus, Markus and Racine, Benjamin and Radeka, Veljko and Ramel, Manon and Ranabhat, Arianna and Rasmussen, Andrew P and Rathfelder, David A and Rawls, Meredith L and Reed, Sophie L and Reil, Kevin A and Reiss, David J and Reuter, Michael A and Ribeiro, Tiago and Rigault, Mickael and Riot, Vincent J and Ritz, Steven M and Rivera, Mario F Rivera and Robertson, Brant E and Roby, William and Rodeghiero, Gabriele and Roodman, Aaron and Rosignoli, Luca and Roucelle, Cecile and Rumore, Matthew R and Russo, Stefano and Rykoff, Eli S and Salnikov, Andrei and Sanchez, Bruno O and Sanmartim, David and Saunders, Clare and Schindler, Rafe H and Schmidt, Samuel J and Sebag, Jacques and Sedaghat, Nima and Selvy, Brian and Valenzuela, Edgard Esteban Sepulveda and Seriche, Gonzalo and Navarrete, Jacqueline C Seron and Noarbe, Ignacio Sevilla and Shugart, Alysha B and Sick, Jonathan and Silva, Cristian and Sims, Mathew C and Singhal, Jaladh and Siruno, Kevin Benjamin and Slater, Colin T and Smart, Brianna M and Snyder, Adam and Soldahl, Christine and Elorriaga, Ioana Sotuela and Stalder, Brian and Stockebrand, Hernan and Strauss, Alan L and Strauss, Michael A and Suberlak, Krzysztof and Sullivan, Ian S and Swinbank, John D and Tapia, Diego and Taranto, Alessio and Taranu, Dan S and Thayer, John Gregg and Thomas, Sandrine and Thornton, Adam J and Tighe, Roberto and Cipriano, Laura Toribio San and Tsai, Te Wei and Tucker, Douglas L and Turri, Max and Tyson, J Anthony and Urbach, Elana K and Utsumi, Yousuke and Van Klaveren, Brian and van Reeven, Wouter and Vaucher, Peter Anthony and Venegas, Paulina and Verma, Aprajita and Villarreal, Antonia Sierra and Voutsinas, Stelios and Walter, Christopher W and Wang, Yuankun David and Waters, Christopher Z and Williams, Christina C and Willman, Beth and Wittgen, Matthias and Vasey, W M Wood and Yang, Wei and Yang, Zhaoyu and Yanny, Brian P and Yoachim, Peter and Zhang, Tianqing and Zhou, Conghao and Zilkova, Danica},
  title         = {The Vera C. Rubin Observatory Data Preview 1},
  year          = {2026},
  month         = mar,
  archiveprefix = {arXiv},
  copyright     = {arXiv.org perpetual, non-exclusive license},
  doi           = {10.71929/RUBIN/2570536},
  eprint        = {2603.23786},
  primaryclass  = {astro-ph.IM},
  publisher     = {NSF-DOE Vera C. Rubin Observatory},
}

@Article{Joseph2016,
  author        = {Joseph, R. and Courbin, F. and Starck, J.-L.},
  journal       = {\aap},
  title         = {Multi-band morpho-Spectral Component Analysis Deblending Tool (MuSCADeT): Deblending colourful objects},
  year          = {2016},
  month         = may,
  pages         = {A2},
  volume        = {589},
  archiveprefix = {arXiv},
  doi           = {10.1051/0004-6361/201527923},
  eid           = {A2},
  eprint        = {1603.00473},
  primaryclass  = {astro-ph.IM},
  url           = {https://ui.adsabs.harvard.edu/abs/2016A&A...589A...2J},
}

@Article{Reiman2019,
  author        = {Reiman, David M. and G{\"o}hre, Brett E.},
  journal       = {\mnras},
  title         = {Deblending galaxy superpositions with branched generative adversarial networks},
  year          = {2019},
  month         = may,
  number        = {2},
  pages         = {2617-2627},
  volume        = {485},
  archiveprefix = {arXiv},
  doi           = {10.1093/mnras/stz575},
  eprint        = {1810.10098},
  primaryclass  = {astro-ph.IM},
  url           = {https://ui.adsabs.harvard.edu/abs/2019MNRAS.485.2617R},
}

@Article{Hemmati2022,
  author        = {Hemmati, Shoubaneh and Huff, Eric and Nayyeri, Hooshang and Fert{\'e}, Agn{\`e}s and Melchior, Peter and Mobasher, Bahram and Rhodes, Jason and Shahidi, Abtin and Teplitz, Harry},
  journal       = {\apj},
  title         = {Deblending Galaxies with Generative Adversarial Networks},
  year          = {2022},
  month         = dec,
  number        = {2},
  pages         = {141},
  volume        = {941},
  archiveprefix = {arXiv},
  doi           = {10.3847/1538-4357/aca1b8},
  eid           = {141},
  eprint        = {2211.04488},
  primaryclass  = {astro-ph.CO},
  url           = {https://ui.adsabs.harvard.edu/abs/2022ApJ...941..141H},
}

@Article{Lanusse2019,
  author        = {{Lanusse}, Francois and {Melchior}, Peter and {Moolekamp}, Fred},
  journal       = {arXiv e-prints},
  title         = {{Hybrid Physical-Deep Learning Model for Astronomical Inverse Problems}},
  year          = {2019},
  month         = dec,
  pages         = {arXiv:1912.03980},
  adsurl        = {https://ui.adsabs.harvard.edu/abs/2019arXiv191203980L},
  archiveprefix = {arXiv},
  doi           = {10.48550/arXiv.1912.03980},
  eid           = {arXiv:1912.03980},
  eprint        = {1912.03980},
  primaryclass  = {astro-ph.IM},
}

@Article{Wang2022,
  author        = {Wang, Hong and Sreejith, Sreevarsha and Slosar, An{\v{z}}e and Lin, Yuewei and Yoo, Shinjae},
  journal       = {\prd},
  title         = {Galaxy deblending using residual dense neural networks},
  year          = {2022},
  month         = sep,
  number        = {6},
  pages         = {063023},
  volume        = {106},
  archiveprefix = {arXiv},
  doi           = {10.1103/PhysRevD.106.063023},
  eid           = {063023},
  eprint        = {2109.09550},
  primaryclass  = {astro-ph.GA},
  url           = {https://ui.adsabs.harvard.edu/abs/2022PhRvD.106f3023W},
}

@Article{Sampson2024,
  author        = {Sampson, M.~L. and Melchior, P. and Ward, C. and Birmingham, S.},
  journal       = {Astronomy and Computing},
  title         = {Score-matching neural networks for improved multi-band source separation},
  year          = {2024},
  month         = oct,
  pages         = {100875},
  volume        = {49},
  archiveprefix = {arXiv},
  doi           = {10.1016/j.ascom.2024.100875},
  eid           = {100875},
  eprint        = {2401.07313},
  primaryclass  = {astro-ph.IM},
  url           = {https://ui.adsabs.harvard.edu/abs/2024A&C....4900875S},
}

@Article{Mendoza2026,
  author        = {Mendoza, Ismael and Hansen, Derek and Liu, Runjing and Zhao, Zhe and Pang, Ziteng and Guinot, Axel and Avestruz, Camille and Regier, Jeffrey and the LSST Dark Energy Science Collaboration},
  journal       = {arXiv e-prints},
  title         = {Simulation-Based Inference for Probabilistic Galaxy Detection and Deblending},
  year          = {2026},
  month         = jan,
  pages         = {arXiv:2601.03422},
  archiveprefix = {arXiv},
  doi           = {10.48550/arXiv.2601.03422},
  eid           = {arXiv:2601.03422},
  eprint        = {2601.03422},
  primaryclass  = {astro-ph.IM},
  url           = {https://ui.adsabs.harvard.edu/abs/2026arXiv260103422M},
}

@Article{Xu2024,
  author        = {Xu, D. and Zhu, Y.},
  journal       = {Astronomy and Computing},
  title         = {Surveying image segmentation approaches in astronomy},
  year          = {2024},
  month         = jul,
  pages         = {100838},
  volume        = {48},
  archiveprefix = {arXiv},
  doi           = {10.1016/j.ascom.2024.100838},
  eid           = {100838},
  eprint        = {2405.14238},
  primaryclass  = {astro-ph.IM},
  url           = {https://ui.adsabs.harvard.edu/abs/2024A&C....4800838X},
}

@Article{Burke2019,
  author        = {Burke, Colin J. and Aleo, Patrick D. and Chen, Yu-Ching and Liu, Xin and Peterson, John R. and Sembroski, Glenn H. and Lin, Joshua Yao-Yu},
  journal       = {\mnras},
  title         = {Deblending and classifying astronomical sources with Mask R-CNN deep learning},
  year          = {2019},
  month         = {Dec},
  number        = {3},
  pages         = {3952-3965},
  volume        = {490},
  archiveprefix = {arXiv},
  doi           = {10.1093/mnras/stz2845},
  eprint        = {1908.02748},
  primaryclass  = {astro-ph.IM},
  url           = {https://ui.adsabs.harvard.edu/abs/2019MNRAS.490.3952B},
}

@Article{Merz2026,
  author   = {Merz, Grant and Zhuang, Ming-Yang and Li, Junyao and Yang, Qian and Shen, Yue and Liu, Xin and Crenshaw, John Franklin},
  journal  = {The Open Journal of Astrophysics},
  title    = {Photometric Redshifts in JWST Deep Fields: A Pixel-Based Alternative with DeepDISC},
  year     = {2026},
  month    = feb,
  pages    = {56099},
  volume   = {9},
  doi      = {10.33232/001c.156099},
  url      = {https://ui.adsabs.harvard.edu/abs/2026OJAp....956099M},
}

@Article{Hausen2020,
  author        = {Hausen, Ryan and Robertson, Brant E.},
  journal       = {\apjs},
  title         = {Morpheus: A Deep Learning Framework for the Pixel-level Analysis of Astronomical Image Data},
  year          = {2020},
  month         = may,
  number        = {1},
  pages         = {20},
  volume        = {248},
  archiveprefix = {arXiv},
  doi           = {10.3847/1538-4365/ab8868},
  eid           = {20},
  eprint        = {1906.11248},
  primaryclass  = {astro-ph.GA},
  url           = {https://ui.adsabs.harvard.edu/abs/2020ApJS..248...20H},
}

@Article{Ward2025,
  author        = {Ward, C. and Melchior, P. and Sampson, M.~L. and Burke, C.~J. and Siegel, J. and Remy, B. and Birmingham, S. and Ramey, E. and Velzen, S. van},
  journal       = {Astronomy and Computing},
  title         = {Disentangling transients and their host galaxies with scarlet2: A framework to forward model multi-epoch imaging},
  year          = {2025},
  month         = apr,
  pages         = {100930},
  volume        = {51},
  archiveprefix = {arXiv},
  doi           = {10.1016/j.ascom.2025.100930},
  eid           = {100930},
  eprint        = {2409.15427},
  primaryclass  = {astro-ph.IM},
  url           = {https://ui.adsabs.harvard.edu/abs/2025A&C....5100930W},
}

@Article{Boucaud2020,
  author        = {Boucaud, Alexandre and Huertas-Company, Marc and Heneka, Caroline and Ishida, Emille E. O. and Sedaghat, Nima and de Souza, Rafael S. and Moews, Ben and Dole, Herv{\'e} and Castellano, Marco and Merlin, Emiliano and Roscani, Valerio and Tramacere, Andrea and Killedar, Madhura and Trindade, Arlindo M. M. and Collaboration COIN},
  journal       = {\mnras},
  title         = {Photometry of high-redshift blended galaxies using deep learning},
  year          = {2020},
  month         = jan,
  number        = {2},
  pages         = {2481-2495},
  volume        = {491},
  archiveprefix = {arXiv},
  doi           = {10.1093/mnras/stz3056},
  eprint        = {1905.01324},
  primaryclass  = {astro-ph.GA},
  url           = {https://ui.adsabs.harvard.edu/abs/2020MNRAS.491.2481B},
}

@article{SEP_2016, doi = {10.21105/joss.00058}, url = {https://doi.org/10.21105/joss.00058}, year = {2016}, publisher = {The Open Journal}, volume = {1}, number = {6}, pages = {58}, author = {Barbary, Kyle}, title = {SEP: Source Extractor as a library}, journal = {Journal of Open Source Software} }

@ARTICLE{desi_legacy,
       author = {{Dey}, Arjun and {Schlegel}, David J. and {Lang}, Dustin and {Blum}, Robert and {Burleigh}, Kaylan and {Fan}, Xiaohui and {Findlay}, Joseph R. and {Finkbeiner}, Doug and {Herrera}, David and {Juneau}, St{\'e}phanie and {Landriau}, Martin and {Levi}, Michael and {McGreer}, Ian and {Meisner}, Aaron and {Myers}, Adam D. and {Moustakas}, John and {Nugent}, Peter and {Patej}, Anna and {Schlafly}, Edward F. and {Walker}, Alistair R. and {Valdes}, Francisco and {Weaver}, Benjamin A. and {Y{\`e}che}, Christophe and {Zou}, Hu and {Zhou}, Xu and {Abareshi}, Behzad and {Abbott}, T.~M.~C. and {Abolfathi}, Bela and {Aguilera}, C. and {Alam}, Shadab and {Allen}, Lori and {Alvarez}, A. and {Annis}, James and {Ansarinejad}, Behzad and {Aubert}, Marie and {Beechert}, Jacqueline and {Bell}, Eric F. and {BenZvi}, Segev Y. and {Beutler}, Florian and {Bielby}, Richard M. and {Bolton}, Adam S. and {Brice{\~n}o}, C{\'e}sar and {Buckley-Geer}, Elizabeth J. and {Butler}, Karen and {Calamida}, Annalisa and {Carlberg}, Raymond G. and {Carter}, Paul and {Casas}, Ricard and {Castander}, Francisco J. and {Choi}, Yumi and {Comparat}, Johan and {Cukanovaite}, Elena and {Delubac}, Timoth{\'e}e and {DeVries}, Kaitlin and {Dey}, Sharmila and {Dhungana}, Govinda and {Dickinson}, Mark and {Ding}, Zhejie and {Donaldson}, John B. and {Duan}, Yutong and {Duckworth}, Christopher J. and {Eftekharzadeh}, Sarah and {Eisenstein}, Daniel J. and {Etourneau}, Thomas and {Fagrelius}, Parker A. and {Farihi}, Jay and {Fitzpatrick}, Mike and {Font-Ribera}, Andreu and {Fulmer}, Leah and {G{\"a}nsicke}, Boris T. and {Gaztanaga}, Enrique and {George}, Koshy and {Gerdes}, David W. and {Gontcho}, Satya Gontcho A. and {Gorgoni}, Claudio and {Green}, Gregory and {Guy}, Julien and {Harmer}, Diane and {Hernandez}, M. and {Honscheid}, Klaus and {Huang}, Lijuan Wendy and {James}, David J. and {Jannuzi}, Buell T. and {Jiang}, Linhua and {Joyce}, Richard and {Karcher}, Armin and {Karkar}, Sonia and {Kehoe}, Robert and {Kneib}, Jean-Paul and {Kueter-Young}, Andrea and {Lan}, Ting-Wen and {Lauer}, Tod R. and {Le Guillou}, Laurent and {Le Van Suu}, Auguste and {Lee}, Jae Hyeon and {Lesser}, Michael and {Perreault Levasseur}, Laurence and {Li}, Ting S. and {Mann}, Justin L. and {Marshall}, Robert and {Mart{\'\i}nez-V{\'a}zquez}, C.~E. and {Martini}, Paul and {du Mas des Bourboux}, H{\'e}lion and {McManus}, Sean and {Meier}, Tobias Gabriel and {M{\'e}nard}, Brice and {Metcalfe}, Nigel and {Mu{\~n}oz-Guti{\'e}rrez}, Andrea and {Najita}, Joan and {Napier}, Kevin and {Narayan}, Gautham and {Newman}, Jeffrey A. and {Nie}, Jundan and {Nord}, Brian and {Norman}, Dara J. and {Olsen}, Knut A.~G. and {Paat}, Anthony and {Palanque-Delabrouille}, Nathalie and {Peng}, Xiyan and {Poppett}, Claire L. and {Poremba}, Megan R. and {Prakash}, Abhishek and {Rabinowitz}, David and {Raichoor}, Anand and {Rezaie}, Mehdi and {Robertson}, A.~N. and {Roe}, Natalie A. and {Ross}, Ashley J. and {Ross}, Nicholas P. and {Rudnick}, Gregory and {Safonova}, Sasha and {Saha}, Abhijit and {S{\'a}nchez}, F. Javier and {Savary}, Elodie and {Schweiker}, Heidi and {Scott}, Adam and {Seo}, Hee-Jong and {Shan}, Huanyuan and {Silva}, David R. and {Slepian}, Zachary and {Soto}, Christian and {Sprayberry}, David and {Staten}, Ryan and {Stillman}, Coley M. and {Stupak}, Robert J. and {Summers}, David L. and {Sien Tie}, Suk and {Tirado}, H. and {Vargas-Maga{\~n}a}, Mariana and {Vivas}, A. Katherina and {Wechsler}, Risa H. and {Williams}, Doug and {Yang}, Jinyi and {Yang}, Qian and {Yapici}, Tolga and {Zaritsky}, Dennis and {Zenteno}, A. and {Zhang}, Kai and {Zhang}, Tianmeng and {Zhou}, Rongpu and {Zhou}, Zhimin},
        title = "{Overview of the DESI Legacy Imaging Surveys}",
      journal = {\aj},
         year = 2019,
        month = may,
       volume = {157},
       number = {5},
          eid = {168},
        pages = {168},
          doi = {10.3847/1538-3881/ab089d},
archivePrefix = {arXiv},
       eprint = {1804.08657},
 primaryClass = {astro-ph.IM},
       adsurl = {https://ui.adsabs.harvard.edu/abs/2019AJ....157..168D}
}

@ARTICLE{pan_starrs,
       author = {{Scolnic}, D.~M. and {Jones}, D.~O. and {Rest}, A. and {Pan}, Y.~C. and {Chornock}, R. and {Foley}, R.~J. and {Huber}, M.~E. and {Kessler}, R. and {Narayan}, G. and {Riess}, A.~G. and {Rodney}, S. and {Berger}, E. and {Brout}, D.~J. and {Challis}, P.~J. and {Drout}, M. and {Finkbeiner}, D. and {Lunnan}, R. and {Kirshner}, R.~P. and {Sanders}, N.~E. and {Schlafly}, E. and {Smartt}, S. and {Stubbs}, C.~W. and {Tonry}, J. and {Wood-Vasey}, W.~M. and {Foley}, M. and {Hand}, J. and {Johnson}, E. and {Burgett}, W.~S. and {Chambers}, K.~C. and {Draper}, P.~W. and {Hodapp}, K.~W. and {Kaiser}, N. and {Kudritzki}, R.~P. and {Magnier}, E.~A. and {Metcalfe}, N. and {Bresolin}, F. and {Gall}, E. and {Kotak}, R. and {McCrum}, M. and {Smith}, K.~W.},
        title = "{The Complete Light-curve Sample of Spectroscopically Confirmed SNe Ia from Pan-STARRS1 and Cosmological Constraints from the Combined Pantheon Sample}",
      journal = {\apj},
         year = 2018,
        month = jun,
       volume = {859},
       number = {2},
          eid = {101},
        pages = {101},
          doi = {10.3847/1538-4357/aab9bb},
archivePrefix = {arXiv},
       eprint = {1710.00845},
 primaryClass = {astro-ph.CO},
       adsurl = {https://ui.adsabs.harvard.edu/abs/2018ApJ...859..101S}
}

@ARTICLE{goods_survey,
       author = {{Giavalisco}, M. and {Ferguson}, H.~C. and {Koekemoer}, A.~M. and {Dickinson}, M. and {Alexander}, D.~M. and {Bauer}, F.~E. and {Bergeron}, J. and {Biagetti}, C. and {Brandt}, W.~N. and {Casertano}, S. and {Cesarsky}, C. and {Chatzichristou}, E. and {Conselice}, C. and {Cristiani}, S. and {Da Costa}, L. and {Dahlen}, T. and {de Mello}, D. and {Eisenhardt}, P. and {Erben}, T. and {Fall}, S.~M. and {Fassnacht}, C. and {Fosbury}, R. and {Fruchter}, A. and {Gardner}, J.~P. and {Grogin}, N. and {Hook}, R.~N. and {Hornschemeier}, A.~E. and {Idzi}, R. and {Jogee}, S. and {Kretchmer}, C. and {Laidler}, V. and {Lee}, K.~S. and {Livio}, M. and {Lucas}, R. and {Madau}, P. and {Mobasher}, B. and {Moustakas}, L.~A. and {Nonino}, M. and {Padovani}, P. and {Papovich}, C. and {Park}, Y. and {Ravindranath}, S. and {Renzini}, A. and {Richardson}, M. and {Riess}, A. and {Rosati}, P. and {Schirmer}, M. and {Schreier}, E. and {Somerville}, R.~S. and {Spinrad}, H. and {Stern}, D. and {Stiavelli}, M. and {Strolger}, L. and {Urry}, C.~M. and {Vandame}, B. and {Williams}, R. and {Wolf}, C.},
        title = "{The Great Observatories Origins Deep Survey: Initial Results from Optical and Near-Infrared Imaging}",
      journal = {\apjl},
         year = 2004,
        month = jan,
       volume = {600},
       number = {2},
        pages = {L93-L98},
          doi = {10.1086/379232},
archivePrefix = {arXiv},
       eprint = {astro-ph/0309105},
 primaryClass = {astro-ph},
       adsurl = {https://ui.adsabs.harvard.edu/abs/2004ApJ...600L..93G}
}

@ARTICLE{astropy_13,
       author = {{Astropy Collaboration} and {Robitaille}, Thomas P. and {Tollerud}, Erik J. and {Greenfield}, Perry and {Droettboom}, Michael and {Bray}, Erik and {Aldcroft}, Tom and {Davis}, Matt and {Ginsburg}, Adam and {Price-Whelan}, Adrian M. and {Kerzendorf}, Wolfgang E. and {Conley}, Alexander and {Crighton}, Neil and {Barbary}, Kyle and {Muna}, Demitri and {Ferguson}, Henry and {Grollier}, Fr{\'e}d{\'e}ric and {Parikh}, Madhura M. and {Nair}, Prasanth H. and {Unther}, Hans M. and {Deil}, Christoph and {Woillez}, Julien and {Conseil}, Simon and {Kramer}, Roban and {Turner}, James E.~H. and {Singer}, Leo and {Fox}, Ryan and {Weaver}, Benjamin A. and {Zabalza}, Victor and {Edwards}, Zachary I. and {Azalee Bostroem}, K. and {Burke}, D.~J. and {Casey}, Andrew R. and {Crawford}, Steven M. and {Dencheva}, Nadia and {Ely}, Justin and {Jenness}, Tim and {Labrie}, Kathleen and {Lim}, Pey Lian and {Pierfederici}, Francesco and {Pontzen}, Andrew and {Ptak}, Andy and {Refsdal}, Brian and {Servillat}, Mathieu and {Streicher}, Ole},
        title = "{Astropy: A community Python package for astronomy}",
      journal = {\aap},
         year = 2013,
        month = oct,
       volume = {558},
          eid = {A33},
        pages = {A33},
          doi = {10.1051/0004-6361/201322068},
archivePrefix = {arXiv},
       eprint = {1307.6212},
 primaryClass = {astro-ph.IM},
       adsurl = {https://ui.adsabs.harvard.edu/abs/2013A&A...558A..33A}
}

@ARTICLE{matplotlib,
  author={Hunter, John D.},
  journal={Computing in Science \& Engineering}, 
  title={Matplotlib: A 2D Graphics Environment}, 
  year={2007},
  volume={9},
  number={3},
  pages={90-95},
  doi={10.1109/MCSE.2007.55}}

@article{numpy,
  title={Array programming with NumPy},
  author={Harris, Charles R and Millman, K Jarrod and Van Der Walt, St{\'e}fan J and Gommers, Ralf and Virtanen, Pauli and Cournapeau, David and Wieser, Eric and Taylor, Julian and Berg, Sebastian and Smith, Nathaniel J and others},
  journal={nature},
  volume={585},
  number={7825},
  pages={357--362},
  year={2020},
  publisher={Nature Publishing Group UK London}
}
